\documentclass[12pt]{iopart}
\usepackage[dvips]{graphicx}
\begin{document}
\title{Exchange bias effect in alloys and compounds}
\author{S Giri, M Patra and S Majumdar}
\ead{sspsg2@iacs.res.in (S Giri)}
\address{Department of Solid State Physics, Indian Association for the Cultivation of Science, Jadavpur, Kolkata 700 032, India}

\begin{abstract}
The phenomenology of exchange bias effects observed in {\it structurally single-phase} alloys and compounds but composed of a variety of coexisting {\it magnetic phases} such as ferromagnetic, antiferromagnetic, ferrimagnetic, spin-glass, cluster-glass, disordered magnetic states are reviewed. The investigations on exchange bias effects are discussed in diverse types of alloys and compounds where qualitative and quantitative aspects of magnetism are focused based on macroscopic experimental tools such as magnetization and magnetoresistance measurements. Here, we focus on improvement of fundamental issues of the exchange bias effects rather than on their technological importance.
   
\end{abstract}


\maketitle


\setcounter{tocdepth}{4}
\tableofcontents

\newpage




\nopagebreak

\vskip 0.5cm

{\centerline{\bf Abbreviations and symbols}}
\vskip 0.3cm  
\noindent Exchange bias \dotfill EB      \\
\noindent Exchange bias field \dotfill $H_E$      \\
\noindent Exchange bias magnetization \dotfill $M_E$      \\
\noindent Cooling field \dotfill $H_{cool}$      \\
\noindent Maximum field applied for recording magnetic hysteresis loop \dotfill $H_{max}$ \\
\noindent Field-cooled \dotfill FC \\
\noindent Zero-field cooled \dotfill ZFC \\
\noindent Saturation magnetization \dotfill $M_S$ \\
\noindent Magnetic field \dotfill $H$  \\
\noindent Temperature \dotfill $T$  \\
\noindent Ferromagnetic \dotfill FM      \\
\noindent Antiferromagnetic \dotfill AFM      \\
\noindent Ferrimagnetic \dotfill FIM      \\
\noindent Pinned ferromagnetic \dotfill PFM      \\
\noindent Frozen ferromagnetic \dotfill FFM      \\
\noindent Spin-glass \dotfill SG      \\
\noindent Cluster-glass \dotfill CG      \\
\noindent Re-entrant spin-glass \dotfill RSG      \\
\noindent Rudermann Kittel Kasuya Yosida \dotfill RKKY      \\
\noindent Resistivity \dotfill $\rho$      \\
\noindent Magnetoresistance $\left(=~\frac{\rho(H)-\rho(0)}{\rho(0)}\right)$
                             \dotfill      MR         \\
\noindent Ferromagnetic Curie temperature \dotfill   $T_C$  \\
\noindent Antiferromagnetic N\'{e}el temperature \dotfill $T_N$ \\


\newpage
\section{Introduction}

The exchange bias effect is an old phenomenon which was discovered in 1956 by Meiklejohn and Bean \cite{meik1}. They observed a shift of the magnetic hysteresis loop along the field axis at low temperature when Co/CoO core/shell nanoparticles were cooled in a static magnetic field \cite{meik1,meik2}. Initially, EB effect was observed in a system composed of ferromagnetic and antiferromagnetic substances where Curie temperature of the FM component is greater than that of the N\'{e}el temperature corresponding to AFM component. When the sample is exposed to a static magnetic field at a temperature ($> T_N$) and cooled through $T_N$ down to low temperature ($\ll T_N$), the FM spins adjacent to AFM spins are coupled to the uncompensated AFM spins. This coupling gives rise to a displacement of magnetic hysteresis loop, a $\sin\theta$ component in the torque curve and  high field rotational hysteresis which are the typical manifestations of EB effect as pointed out by Meiklejhon in a review \cite{meik3}.

After the discovery of EB effect in FM Co/AFM CoO nanostructures the EB has been observed in diverse combinations between FM, AFM, canted AFM, FIM, SG and disordered magnetic components which have been reviewed by different groups \cite{wohl,schm,lub,jacob,kou,yel,march,nogues1,berk,stamp,kiwi,binekREV,coeh,bobo,nogues2,nogues3,iglesias,gra}. Recently, the EB effects were also reported even in a combination of  hard FM and soft FM layered systems either directly exchange coupling  \cite{full,dume,gubb,sort} or coupling through a nonmagnetic interlayer between them \cite{binek-FM-FM,berg,zha}. In the latter case, coupling was suggested to arise from the interplay between dipolar and RKKY interactions between the layers. It is important to point out that measurements of minor loops of the soft FM component were required to investigate  exchange bias in the above two cases \cite{berg_minor}. Investigations on the EB effect have been explored mainly in layered structures and nanoparticles having core-shell nanostructures, addressing different issues on EB phenomenon. Because of the complex mechanisms at different kinds of interfaces, often theoretical interpretations, which are mainly based on macroscopic or phenomenological models, are inadequate. Recently, Monte Carlo simulations or micromagnetic approach \cite{monte-fid,monte-ig1,monte-ig2,monte-hu,monte-trohi1,monte-trohi2,monte-trohi3,monte-trohi4} have been used to gain deeper insight into microscopic origin of EB effect. However, in most of the cases theories were limited to specific problems where intricate structures at the interfaces were simplified. Until now, EB effects have been exploited in several technological applications such as read head of recording devices \cite{rec-tsa1,rec-tsa2,rec-ton1,rec-gan1,rec-lin1,rec-son1,rec-liu1,rec-mac1,rec-doe1,rec-mat1,rec-zha1,rec-ara1,rec-tan1,rec-zha2,rec-hon1,rec-sto1}, magnetoresistive random access memories (MRRAM) \cite{mr-lea1,mr-cro1,mr-kat1,mr-weg,mr-koc,mr-rus1,mr-rus2,mr-sun1,mr-poh1,mr-ric1,mr-smi1,mr-osh1,mr-kak1,mr-li1,mr-liu1,mr-par1,mr-teh1,mr-wor1,mr-wor2,mr-liu2,mr-zhe1,mr-zhe2} and it has been proposed for the technological applications in stabilizing magnetization of superparamagnetic nanoparticles \cite{nog-wel1,nog-tho1,nog-sku1,nog-nog1} or to improve coercivity and energy product of the permanent magnets  \cite{sort1_FM-AFM,sort2_FM-AFM,sort3_FM-AFM,sort4_FM-AFM}.

The EB effects have been observed in numerous morphologies at the artificial interfaces viz., layer structures \cite{nogues1,berk,nogues2}, core-shell nanostructures \cite{iglesias} or irregular metal and metal oxide nanostructures \cite{thakur1,das}, FM nanoparticles embedded in AFM matrix \cite{nogues3,nog-sku1,sort1_FM-AFM,sort2_FM-AFM,sort3_FM-AFM,sort4_FM-AFM,misra,dom} which have been focused to develop advanced materials for the applications and to understand the complex EB phenomenology \cite{nogues1,berk,stamp,kiwi}. However, investigations on the EB effects in structurally single-phase alloys and compounds have been far less. In fact, already in late fifties, Kouvel and coworkers reported extensively on the  evidences of EB effects in different of binary alloys involving disordered magnetic states \cite{kou1}, SG or CG phases \cite{kou2,kou3} and coexisting FM/AFM interactions \cite{kou4,kou5,kou6}. After a long gap, reports on EB effects on structurally single-phase alloys and compounds together with the nanoparticles having FM/AFM/FIM core and disordered magnetic/SG/SG-like shell structures can be found in the literatures (see Refs. [19] and [21]). Recently, EB effects were also reported in a charge-ordered manganite, Pr$_{1/3}$Ca$_{2/3}$MnO$_3$ by Niebieskikwiat and Salamon where EB was attributed to the spontaneous interfaces between short range FM droplets or clusters embedded in the AFM matrix \cite{nieb}. Although few reports of EB in structurally single phase was found including detailed investigation in binary alloys by Kouvel, this report has attracted a renewed attention for investigating EB effects in alloys and compounds having single-phase crystal structure. 

To date, a fairly substantial volume of works have been reported focusing on different aspects of EB in alloys, intermetallic and oxide compounds. In few of the cases, investigations on EB effects using macroscopic experimental approaches such as magnetization or magnetoresistance have been exploited for understanding qualitative and quantitative aspects of nanoscale magnetic phase separations in structurally single-phase alloys and compounds. Herein, we are motivated in reviewing the EB phenomena in single-phase alloys, intermetallic and oxide compounds. The present review has been organised in the following manner. After this current section (section 1) EB  phenomenology is briefly discussed in section 2. In section 3 experimental results of EB effect in alloys and intermetallic compounds are reviewed. In section 4 oxide materials are discussed in the context of EB effect. The EB effect found in structurally single-phase fine particles having magnetic core-shell phase coexistence is reviewed in section 5. Results of EB observed in  magnetoresistance is incorporated in section 6. Finally, evaluation of reported experimental results and general concluding remarks, focusing advancement of fundamental understanding and areas needed to be improved are briefly highlighted in section 7.

\section{Phenomenology of exchange bias}

\begin{figure}[t]
\centering
\includegraphics[width = 9 cm]{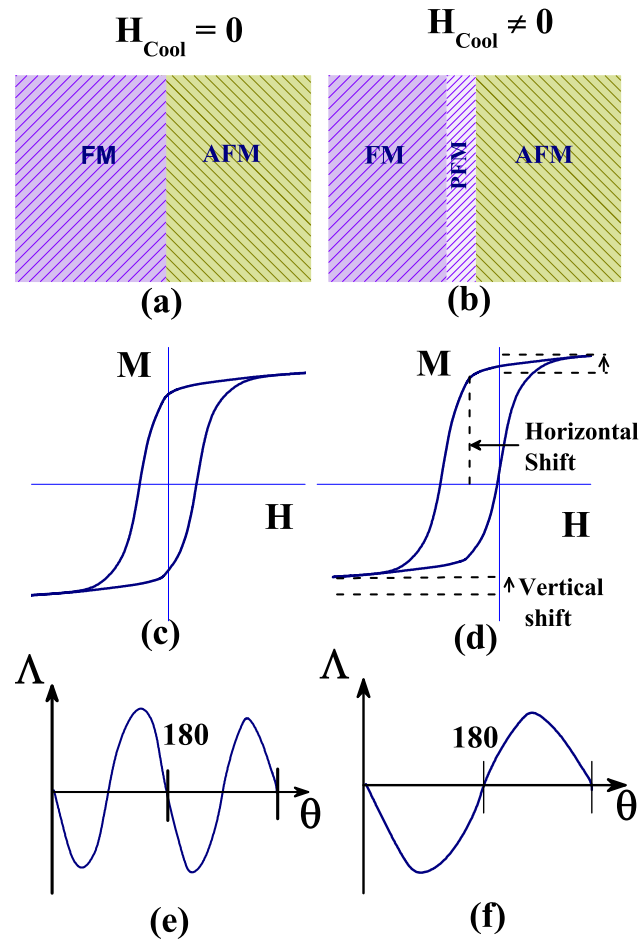}
\caption {Schematic representation of EB effect due to field cooling in  (a) bilayer FM/AFM structure and (b) bilayer structure with appearance of a new pinned ferromagnetic (PFM) layer at the interface; (c) no shift and (d)  loop shifts; (e) uniaxial and (f) unidirectional anisotropies, for cooling the sample in zero-field and a static magnetic field, respectively.}
\end{figure}

\subsection{Basic phenomenology}

\begin{figure}[t]
\centering
\includegraphics[width = 9 cm]{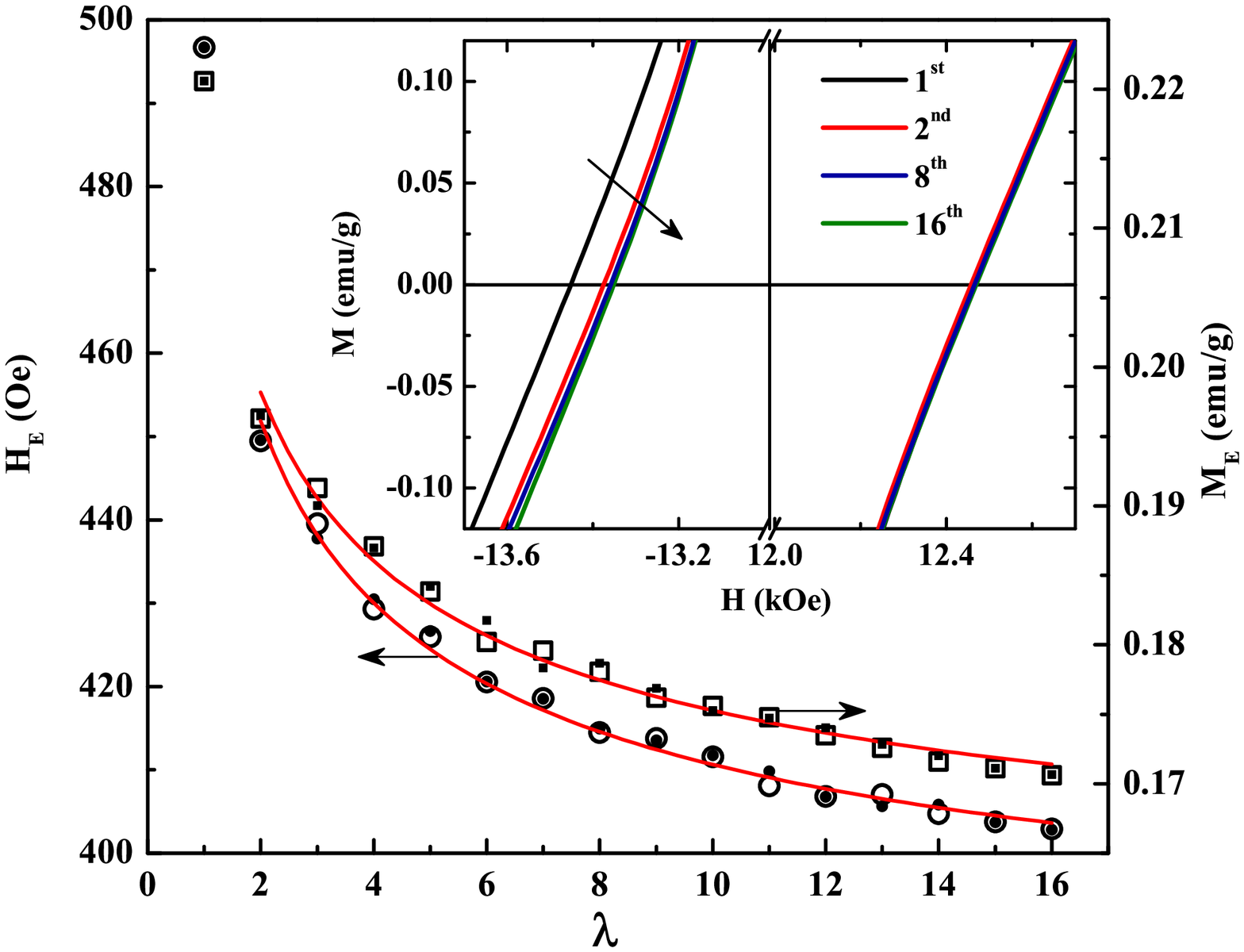}
\caption {Decrease of $H_E$ and $M_E$ with consecutive number of cycles ($\lambda$) of the
loop measurement for Nd$_{0.8}$Sr$_{0.2}$CoO$_3$ exhibiting the training effect. The filled symbols provide the values of $H_E$ and $M_E$ at $\lambda$ obtained from least square fits using equation (2) while continuous curves display the fits using equation (1) for $\lambda \geq$ 2. The
central part of 1st, 2nd, 8th and 16th loops are shown in the inset (The text used in the inset are newly added from the original figure for better clarification), where arrow indicates increasing direction of $\lambda$ [reprinted from \cite{patraJPCM1}].}
\end{figure}

The phenomenology of EB effect has been described explicitly in different reviews \cite{wohl,schm,lub,jacob,kou,yel,march,nogues1,berk,stamp,kiwi,binekREV,coeh,bobo,nogues2,nogues3,iglesias,gra}. Most of those reviews are mainly focused on the experimental evidences through the manifestation of displacement of magnetic hysteresis loop resulting from the  pinning effect at the interface between soft and hard magnetic substances. An example of displaced hysteresis loop is shown in figure 1(d) when the sample comprised of soft and hard magnetic substances is exposed to a static magnetic field during cooling process. As noted in figure 1(d) displacement of magnetic hysteresis loop can be manifested through the horizontal and vertical shifts. The horizontal shift provides $H_E$ whereas $M_E$ is typically estimated from the vertical shift of hysteresis loop at saturation magnetization [displayed in figure 1(d)]. The shift is absent while the sample is cooled in zero-field as displayed in figure 1(c). The shift along the field axis is  typically negative for positive cooling field and it is positive for negative cooling field. For simplicity, let us consider a bilayer system composed of FM and AFM substances where FM and AFM substances represent  soft and hard magnetic substances. The simplified FM/AFM bilayer structure is shown in figure 1(a). When the sample is cooled in a static field through $T_N$ from a temperature, $T_N < T < T_C$ down to low temperature ($T \ll T_N$), a new layer appears at the interface displayed in figure 1(b). This new layer is called as the pinned FM layer consisting of pinned FM spins which gives rise to the displacement of magnetic hysteresis loop. The FM spins adjacent to the interface are pinned by AFM spins in FC mode. The pinned FM layer has an unidirectional anisotropy. The unidirectional anisotropy direction is set by direction of cooling field and is usually set along uniaxial direction of the FM component. Thus, the pinned FM layer provides a new type of anisotropy which has a $K_{ud} \cos(\theta)$ angular dependence rather than the uniaxial anisotropy having $K_{ua} \sin(2\theta)$ dependence. The $\theta$, $K_{ud}$ and $K_{ua}$ are angle between direction of applied magnetic field and anisotropy direction, unidirectional anisotropy constant of pinned FM spins and uniaxial anisotropy constant of the FM component, respectively. Figures 1(e) and 1(f) illustrate typical example of the plots of torque magnetometry, $\Lambda$ with $\theta$ for both the cases where unidirectional anisotropy has a single minimum energy state in contrast to two minima found in case of uniaxial anisotropy.

\subsection{Training effect}

\begin{figure}[t]
\centering
\includegraphics[width = 8 cm]{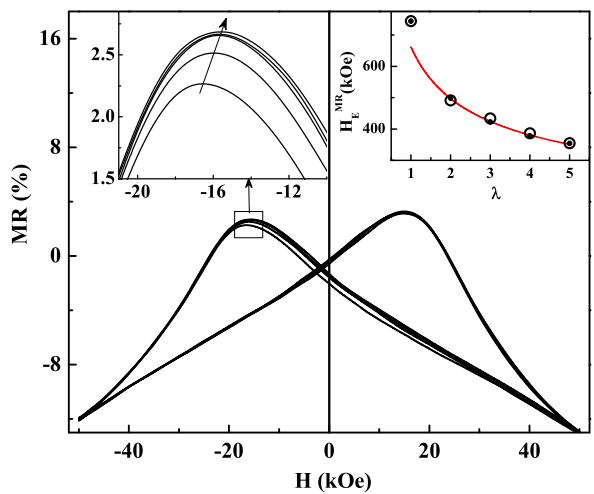}
\caption {Training effect is shown in the MR-$H$ curves up to 5
successive cycles ($\lambda$ = 5) in Nd$_{0.84}$Sr$_{0.16}$CoO$_3$. (Left inset) highlights the shift of peak position with increasing $\lambda$ where arrow indicates increasing
direction of $\lambda$. (Right inset) Plot of $H_E$ with $\lambda$ where solid straight line exhibits the fit by a power law given in equation (1) (In the original figure $H_E$ was plotted with $\lambda^{-0.5}$). Open circles are experimental data while small filled symbols represent the data obtained from least square fit described in equation (2) [reprinted from \cite{patraJPCM2}].}
\end{figure}

Training effect is one of the important experimental evidences which is usually observed in EB systems. Difference between subsequent magnetization reversal loops is called the training effect. This is investigated by successive field cycling at low temperature after the field cooling. The evidence of training effect was first reported in a thin film \cite{pacc}. They further proposed that decrease of $H_E$ or $M_E$ satisfies following empirical formula 
\begin{equation}
H_E - H_E^{\infty} \propto \lambda^{-1/2} 
\end{equation}
where $\lambda$ is the loop index number and $H_E^{\infty}$ the value at $\lambda = \infty$. Two kinds of training effects have been reported: 1) decrease of EB effect for $\lambda <$ 2 \cite{hoff} and 2) decrease for $\lambda \geq$ 2 \cite{nogues2,iglesias}. Equation (1) can be derived in a series expansion approach which satisfies the second type of training effect i.e. for $\lambda \geq$ 2. The training effect in magnetic hysteresis loop \cite{patraJPCM1} and MR curve \cite{patraJPCM2} has been shown for Nd$_{0.8}$Sr$_{0.2}$CoO$_3$ in figures 2 and 3, respectively. The decreases in $H_E$ are highlighted by the arrows in the insets of the figures. The continuous curves in figure 2 and right inset of figure 3 display fits of the experimental data with equation (1) for $\lambda \geq$ 2. The more generalized interpretation of  training effect was recently proposed by Binek with the help of a discretized  Landau-Khalatnikov equation \cite{LK} where continuous time parameter was replaced by $\lambda$ and arrived at a general recursive formula for $H_E$ or $M_E$ \cite{binek}
\begin{equation}
H_E (\lambda + 1) - H_E (\lambda) = -\gamma [H_E (\lambda) - H_E (\lambda = \infty)]^3 
\end{equation}
where $\gamma$ is a constant. It is to be noted that equation (1) can be derived from the series expansion of equation (2) where equation (1) is limiting case of equation (2) and can fit the training effect for $\lambda \geq$ 2. On the other hand, equation (2) is a more general form which satisfies the training effect for all $\lambda$. The values of $H_E$ or $M_E$ obtained from the least square fit with $\lambda$ satisfactorily interpret the experimental data using equation (2) for different cases such as layered structures \cite{binek1,binek2,binek3,binek4,binek5,guh,yang,ang,luo1,yangpy}, nanocomposites \cite{tian}, bulk Heusler alloys \cite{wang-EB-JAP-TE,nayak-EB-JPD-TE} and oxide compounds \cite{nieb,patraJPCM1,tang-EB-PRB,BKC-EB-PRB,patraSSC,yuaLSCO,patraJAP}. Equation (2) can also satisfactorily interpret the shift of MR curves, exhibiting EB effect \cite{patraJPCM2,vent,patraEPL}. As seen in figure 2 and right inset of figure 3, the  values (filled symbols) as obtained from the fit using equation (2) can satisfactorily match with all the experimental data. Recently, the exponential dependence of training effect with $\lambda$ has been demonstrated in a NiFe/IrMn bilayer \cite{mishra} and Pt/Co/Pt/IrMn multilayers films \cite{shi}. This training effect was interpreted in terms of metastable magnetic disorder at the magnetically frustrated interface during magnetization reversal process. It is worthwhile to note that the fits using equation (2) and exponential dependence are virtually indistinguishable where exponential bahaviour has more free parameters than equation (2). 

\subsection{Minor loop effect}

\begin{figure}[t]
\centering
\includegraphics[width = 7.5 cm]{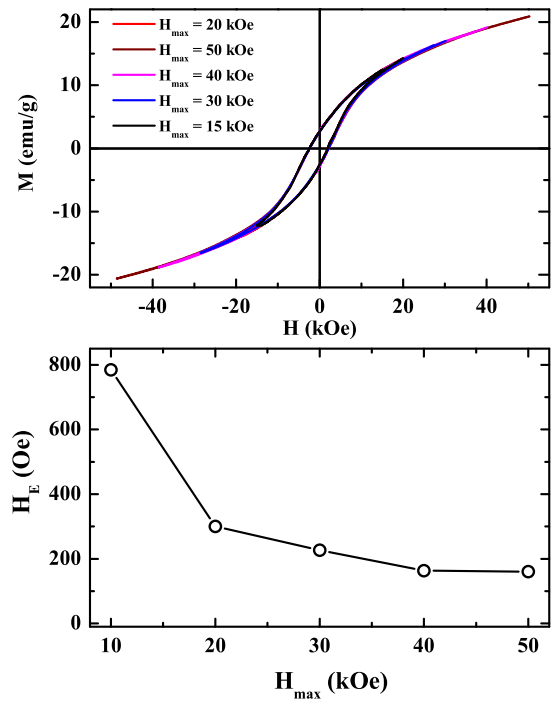}
\caption {Top panel: field-cooled hysteresis loops at 5 K for different $H_{max}$. Bottom panel: plot of $H_E$ against $H_{max}$ at 5 K for LaMn$_{0.7}$Fe$_{0.3}$O$_3$ [reprinted from \cite{giriREP}].}
\end{figure}

In case of nanocrystalline compounds, alloys and oxide materials magnetization at high magnetic field (viz., 50 kOe) does not often saturate. In particular, materials involving  disordered magnetic and/or glassy magnetic phases or canted spin configuration or a system with large anisotropy do not show saturating trend even for $H >$ 50 kOe. Thus, proper choice of maximum field applied for recording  magnetic hysteresis loop, $H_{max}$ is crucial for investigating EB effect. Because small $H_{max}$ may lead to the displacement of magnetic hysteresis loop even for FM and glassy magnetic substances attributed to the irreversible magnetization processes \cite{kle,gesh1,gesh2,luoRES,gesh3}. In order to avoid minor loop effect $H_{max}$ may be chosen such that $H_{max} > H_{A}$ where $H_{A}$ is anisotropy field of the system \cite{gesh1,gesh4,gesh5,gesh6,giriREP}. In order to avoid the overestimated value of EB field  and magnetization, the plots of $H_E$ and $M_E$ with $H_{max}$ are found in few literatures \cite{tang-EB-PRB,yuaLSCO,gesh1,luoRES,giriREP,noguesJACS,luo-LBCO-APL,pi-SRO-APL,yan-CuZnCr2O4-JAP,yan-ZnCr2O4-JPCM,ang-SrPrCoO4-JAP,xu-MR-EB-JPC,liu-NCMO-JPD}. In the plots $H_E$  and $M_E$ decrease with increasing $H_{max}$ and the values stabilize at higher $H_{max}$ \cite{tang-EB-PRB,yuaLSCO,giriREP,noguesJACS,xu-MR-EB-JPC,liu-NCMO-JPD}. An example of $H_{max}$ dependence of $H_E$ is displayed in figure 4 for LaMn$_{0.7}$Fe$_{0.3}$O$_3$ \cite{giriREP}. Magnetic hysteresis loops measured in between different $\pm H_{max}$ are displayed in figure 4(a). The plot of $H_E$ with $H_{max}$ is shown in figure 4(b) which clearly demonstrates that the value of  $H_E$ stabilizes for $H_{max} \geq$ 40 kOe. Nogues and his coworkers reported that the vertical shift vanished whereas the horizontal shift displayed a stabilized value of $H_E$ for large $H_{max}$ at 70 kOe for MnO/Mn$_3$O$_4$ nanoparticles having core-shell structures \cite{noguesJACS}. In few examples EB effect vanished at high enough $H_{max}$, although interpretation of the results were different  \cite{gesh1,luoRES,luo-LBCO-APL,pi-SRO-APL,yan-CuZnCr2O4-JAP,yan-ZnCr2O4-JPCM,ang-SrPrCoO4-JAP}.

\section{Exchange bias effect in alloys and intermetallic compounds}

The conduction electrons in magnetic alloys and intermetallic compounds lead to stronger and long-range indirect-exchange interaction which is named RKKY interaction. The Hamiltonian involved with the RKKY interaction is $H = J(r) {\rm \overline{S}}_i \cdot {\rm \overline{S}}_j$. When the magnetic impurities or localized moments embed in a sea of conduction electron, it cause a damped oscillation in the susceptibility of conduction electrons and thereby, coupling between ${\rm \overline{S}}_i$ and ${\rm \overline{S}}_j$ is given by
\begin{equation}
J(r) = 6\pi ZJ^{2}N(E_{F})\left[\frac{\sin(2{\rm \overline{k}}_{F}r)}{(2{\rm \overline{k}}_{F}r)^{4}} - \frac{\cos(2{\rm \overline{k}}_{F}r)}{(2{\rm \overline{k}}_{F}r)^{3}}\right].
\end{equation} 
In the above expression $Z$ is the number of conduction electrons per atom, $J$ the $s-d$ exchange constant, $N(E_F)$ the density of states at Fermi level, ${\rm \overline{k}}_{F}$ the Fermi wave vector and $r$ the distance between two magnetic impurities. Thus, sign of impurity coupling (positive for FM and negative for AFM interactions) varies with distance ($r$) giving rise to variety of magnetic ground states from Kondo regime, SG, cluster SG, RSG and long range magnetic ordering with increasing concentration of the magnetic impurity \cite{mydosh}. Depending on variations of distances between impurity states, coexisting magnetic phases are quite frequently observed in variety of diluted magnetic alloys. 
Thus, magnetic alloys having coexisting magnetic phases are one of the promising candidates for investigating EB effect. 
Until now, EB effect has been reported in Laves phase intermetallic compounds, binary alloys and few magnetic shape memory alloys. The trend of experimental experimental results  triggers the renewed attention of investigating EB effect in variety of alloys comprised of various coexisting magnetic phases that have well defined ground states. In the next section EB effect in alloys and intermetallic compounds will be reviewed in three following subsections.

\subsection{Laves phase intermetallic compounds and alloys}

\begin{figure}[t]
\centering
\includegraphics[width = 7.5 cm]{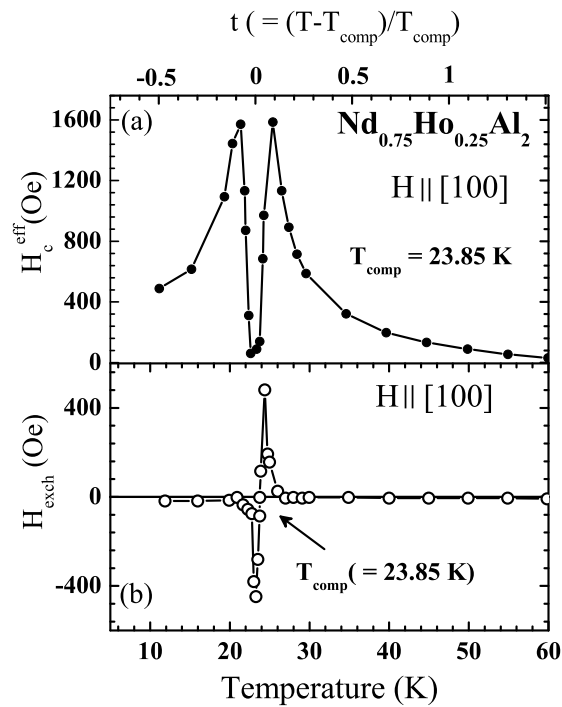}
\caption{Temperature dependences of (a) effective coercive field ($H_{C}^{eff}$) and  (b) EB  field ($H_{exch}$) in single crystalline Nd$_{0.75}$Ho$_{0.25}$Al$_2$ for $H \|$ to  [100] direction [reprinted with permission from \cite{EB-thamiz-EPL}].}
\end{figure}

First signature of EB effect was reported in a Laves phase intermetallic compound, UMn$_2$ \cite{lin-UMn2}. When the sample was cooled in a static field, an asymmetry in magnetic hysteresis loop was noticed. This asymmetry was proposed to be a manifestation of EB effect. UMn$_2$ was described as an AFM compound associated with a small parasitic FM component which was attributed to uncompensated sublattice moment or uncompensated domain wall magnetization, although absence of long range magnetic ordering was confirmed recently by $^{55}$Mn NMR and NQR investigations \cite{giri-NMR-NQR}. Very recently, an interesting observation of EB effect has also been reported for R$_{1-x}$Gd$_x$Al$_2$ \cite{EB-thamiz-TOM,EB-thamiz-EPL} and Sm$_{0.975}$Gd$_{0.025}$Cu$_4$Pd \cite{grover1} in the vicinity of compensation temperature, at which ZFC magnetization changed its sign. Analogous to that observed in sign change of magnetization, change in sign of $H_E$ was noticed close to compensation temperature in polycrystalline,  Sm$_{0.98}$Gd$_{0.02}$Al$_2$, polycrystalline, Pr$_{0.8}$Gd$_{0.2}$Al$_2$ and single crystalline, Nd$_{0.75}$Gd$_{0.25}$Al$_2$. An example of temperature dependence of EB field ($H_{exch}$) and corresponding effective coercivity ($H_{C}^{eff}$) are displayed in figure 5 for single crystalline, Nd$_{0.75}$Gd$_{0.25}$Al$_2$ \cite{EB-thamiz-EPL}. Below compensation temperature $H_E$ almost vanished for Sm$_{0.98}$Gd$_{0.02}$Al$_2$ and Nd$_{0.75}$Gd$_{0.25}$Al$_2$ whereas a small value of $H_E$ was observed in Pr$_{0.8}$Gd$_{0.2}$Al$_2$, exhibiting increasing trend of $H_E$ with decreasing temperature. Above a threshold field, the conduction electron polarization was also found to reverse its sign at the compensation temperature. The authors suggested that this effect was correlated with the observed phase reversal in $H_E$. Here, interesting temperature dependence of $H_E$ probes insight into the microscopic views of coexisting magnetic structures in the vicinity of compensation temperature of these alloys. 
Similar to that observed in those Laves phase compounds, rare earth based intermetallics, SmScGe and NdScGe also exhibited EB effect near-zero net
magnetization with substitutions of 6-9 atomic \% of Nd and 25 atomic \% of Gd, respectively \cite{EB-kulk-JPD}. These magnetically ordered materials with 'no-net' magnetization and appreciable conduction electron polarization displayed signature of $H_E$ which could be tuned by the substitution. Recently, Kulkarni {\it et al}. has demonstrated variety of EB phenomenology in FM Sm intermetallic compounds viz., SmCu$_4$Pd, SmPtZn, SmCd, SmZn, SmScGe and SmAl$_2$ having different crystallographic structures \cite{grover1}. They suggested that EB was ascribed to compensation between local 4$f$-orbital and 4$f$-spin contribution of the magnetic moment of Sm$^{3+}$. 

Signature of EB was observed in amorphous ErCo$_2$ film due to field cooling. The EB effect was found to increase with decreasing temperature  down to 4 K, which started from 400 K being the highest available temperature  \cite{EB-oner-PhysicaB}. The EB was proposed to arise from exchange interaction at the microscopic interfaces between antiferromagnetically correlated subdomains and  ferromagnetically correlated subdomains of Er atoms. Magnetic hysteresis measured in a bulk amorphous alloy with composition  Nd$_{60}$Fe$_{30}$Al$_{10}$ cooled in an external field of 100 kOe revealed existence of strong unidirectional anisotropy over a wide temperature range \cite{EB-hong-JMMM}. The $H_E$ was considerably high ($\sim$ 8 kOe) at 4.2 K. The existence of this anisotropy was suggested due to magnetic ordering having two secondary antiferromagnetic phases.

\subsection{Binary alloys}

Kouvel and his co-workers extensively reported the evidences of exchange anisotropy in few binary alloys such as Ni-Mn \cite{kou1,kou3,kou5,kou6,kou7}, Cu-Mn \cite{kou7}, Ag-Mn \cite{kou2}, Co-Mn \cite{kou4} and Fe-Mn \cite{kou5} alloys where some of them have classically well-defined magnetic ground states. Kouvel summarized his works in a review where he demonstrated that exchange anisotropy in variety of binary alloys was attributed to coexistence of FM and AFM or SG phases \cite{kou}. 
The most extensively studied materials are solid solution of Ni-Mn alloys where magnetic ground state as well as exchange anisotropy was changed systematically with increasing Mn concentration. Disordered polycrystalline Ni-Mn alloys of about 20, 25 and 30 atomic percent Mn were cooled from 300 K down to 1.8 K in a magnetic field and their magnetic hysteresis loops were found to be displaced from the origin. The displacement of the loops decreased  monotonically and vanished at around $\sim$ 25, $\sim$ 35 and $\sim$ 75 K for 20, 25 and 30 atomic percent Mn, respectively. Torque magnetometry  exhibited the unidirectional anisotropy. The evidence of EB anisotropy was also noticed for higher Mn concentration with composition Ni$_3$Mn \cite{kou7}. When nickel in composition, Ni$_3$Mn was partially replaced by Fe, a strong exchange anisotropy was observed in alloys for less than about 50 atomic percent Fe \cite{kou5}.  At higher iron concentrations exchange anisotropy disappeared and it was a simple antiferromagnet. They further proposed that Fe-Ni exchange coupling was strongly FM while  both Fe-Mn and Fe-Fe interactions were antiferromagnetic. 

Similar to that observed in Ni-Mn alloys, Co-Mn alloys of about 25, 30 and 35 atomic percent  Mn exhibited signature of exchange anisotropy \cite{kou4}. This behaviour was suggested to arise from exchange anisotropy between regions having FM and AFM spin alignments. It was shown that this result combined with statistical compositional fluctuations inherent to a disordered alloy provided a plausible model of exchange anisotropy in these materials. A significant evidence of EB anisotropy was reported in Al-Fe alloys with 50 \% and 30 \% of Al which was interpreted due to coexistence of ferromagnetism and antiferromagnetism in Al-Fe alloys \cite{kou8}. Kouvel proposed that anisotropy was complicated in nature which was different from the simplified model proposed in Co/CoO nanostructures.     

Kouvel further demonstrated signature of EB effect in classical SG systems such as Cu-Mn and Ag-Mn alloys \cite{kou,kou2}. Magnetic hysteresis loops were  displaced from their symmetrical positions around the origin when alloys having 5-30 \%  Mn in Cu and 10-25 \% Mn in Ag were cooled in a static magnetic field. The displacement of hysteresis loop was found to decrease with increasing temperature and its disappearance was found close to the transition temperature. The displacement in magnetic hysteresis loops was described in terms of EB effect. The coexistence of AFM and FM interactions between Mn atoms having different phase separation between them  was proposed to interpret EB effect. 

\subsection{Heusler alloys}

Magnetic Heusler alloys with composition, X$_2$YZ having face-centered cubic crystal structure undergoing a martensitic transition display variety of interesting functional properties \cite{manosa1,manosa2}. These effects are commonly interpreted as a consequence of strong coupling between structural and magnetic properties. Associated with this first-order magnetostructural transition these materials also display magnetic shape-memory properties. First example of magnetic shape-memory properties was found in Ni$_2$MnGa single crystal \cite{ulla}. The details of magnetostructural properties in magnetic shape memory alloys are beyond the scope of this review. Herein, we focus on observation of EB effect in few  magnetic shape memory alloys. 

\subsubsection{Cu-Mn-Al alloy}

The alloy with composition Cu$_{44.7}$Mn$_{20.6}$Al$_{37.7}$ ranging in between Heusler phase and $\kappa$ phase first exhibited signature of EB anisotropy \cite{koga-EB-JPCS}. The authors suggested that high coercive force, displaced and constricted hysteresis loops observed in this alloy were attributed to exchange anisotropy interactions between FM domains of Heusler phase and AFM Mn-rich domains.

\subsubsection{Ni-Mn-Sb alloys}
The evidence of EB effect was recently reported by Khan {\it et al}. in bulk polycrystalline Ni$_{50}$Mn$_{25+x}$Sb$_{25-x}$ shape memory alloys \cite{khan-EB-APL}. The horizontal shifts in hysteresis loops up to $\sim$ 248 Oe were observed for cooling the sample in 50 kOe magnetic field at low temperature. The observed EB phenomenon was attributed to coexistence of AFM and FM exchange interactions in the system. The results provided direct evidence of coexisting magnetic phases in polycrystalline Ni$_{50}$Mn$_{25+x}$Sb$_{25-x}$ from EB effect at low temperature. It was noticed that $H_E$ increased from 174 to 248 Oe with increasing Mn concentration ranging  $x$ from 12 to 13.5. The detailed composition dependent EB effect and magnetostructural phase diagram were proposed in Ni-Mn-Sb alloys where EB was reported for 10 $\leq x \leq$ 16 in Ni$_{50}$Mn$_{25+x}$Sb$_{25-x}$ \cite{khan-EB-JPCM}. They further demonstrated that temperature, at which EB effect vanished, was found to increase with $x$. In the above composition EB effect ascribed to coexistence of FM and AFM interactions below EB blocking temperature at 100 K was reported at $x$ = 13 \cite{rao-EB-JPD}. The EB blocking temperature was defined by a temperature, above which $H_E$ vanished. The strong EB effect has also been reported at $x$ = 13 in Ni$_{50}$Mn$_{25+x}$Sb$_{25-x}$ as a consequence of Co substitution in Ni-site \cite{nayak-EB-JPD-TE}. Large EB in Ni$_{50-x}$Co$_{x}$Mn$_{38}$Sb$_{12}$ for $x$ = 0, 2, 3, 4, 5 was  attributed to coexistence of FM and AFM phases in the martensitic phase. The $H_E$ was found to increase with increasing $x$. The maximum value of $H_E$ was reported to be $\sim$ 480 Oe at $T$ = 3 K for $x$ = 5  after cooling in 50 kOe which is the highest value reported so far in any Heusler alloys. The increase of AFM coupling resulting from Co substitution was suggested to be responsible for enhancement of EB. They further suggested that phase coexistence appeared due to supercooling of high temperature ferromagnetism and predominant AFM components in the martensitic phase. Training effect was observed which could be satisfactorily analyzed by the empirical formula given in equation (1). Note that $H_E$ involved with the training effect could be analyzed for all $\lambda$ (index number) values unlike other observations where equation (1) typically satisfies $H_E$ with $\lambda$ plots for $\lambda \geq$ 2 \cite{nogues2,iglesias}. 

\subsubsection{Ni-Mn-In alloys}
 
The EB was observed in bulk polycrystalline Ni-Mn-In alloy with composition Ni$_{49.5}$Mn$_{34.5}$In$_{16}$ where FM and AFM phases coexist in the martensitic state \cite{wang-EB-JAP-TE}. The $H_E$ and $H_C$ were strongly dependent on temperature. Training effect was investigated which was found to be very small. The mechanism of training effect in this alloy was attributed to depinning of uncompensated AFM spins. They suggested that FM and AFM domains coupled at the interfaces resulted in the EB effect. The signature of EB was also observed in slightly different composition with Ni$_{50}$Mn$_{35}$In$_{15}$  which was suggested due to coexistence of FM and AFM interactions \cite{jing-EB-JALCOM}. The $H_E$ vanished above 120 K, close to temperature, around which a distinct peak was noticed in the ZFC magnetization. The EB effect has been investigated in bulk Heusler alloys,  Ni$_{50}$Mn$_{50-x}$In$_x$ with  14.5 $\leq x \leq$ 15.2 \cite{pathak-EB-JMMM}. The $H_E$  was found to increase sharply with $x$. It showed almost unchanged value of $H_E$ in the range 14.8 $< x <$ 15.2 and then it showed a sharp decrease at $x$ = 15.2. The EB effect has also been investigated in Ni$_{50}$Mn$_{35}$In$_{15-y}$Si$_{y}$ alloys with $y$ = 0, 1, 2, 3 and 4 where the effect of Si substitution on EB phenomenon was probed \cite{pathak-EB-IEEE-TOM}. The $H_E$ was increased with Si substitution and maximum $H_E$  was observed to be $\sim$ 170 Oe for $x$ = 4 at 5 K for $H_{cool}$ at 50 kOe.

\begin{figure}[t]
\centering
\includegraphics[width = 7.5 cm]{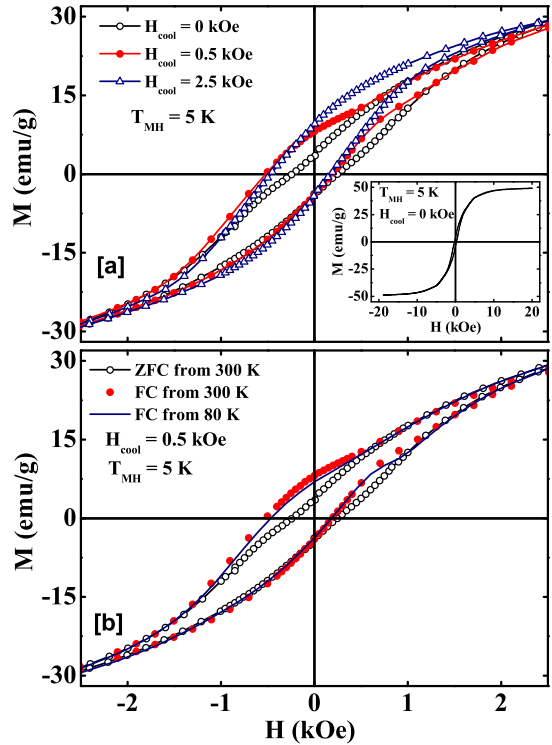}
\caption {(a) Isothermal $M-H$ loops at 5 K after the sample being cooled in $H$ = 0, 0.5 and 2.5 kOe from 300 K for Ni$_{50}$Mn$_{34}$Sn$_{16}$. (b) Isothermal magnetization loops at 5 K after the sample being cooled in zero field, in 0.5 kOe from 300 K and in 0.5 kOe from 80 K. The magnetization loops were recorded by varying field between $\pm$ 20 kOe while data are shown here between $\pm$ 2.5 kOe for the clarity. A full loop on the zero-field-cooled state is shown in the inset of (a). [reprinted from \cite{chat-EB-PRB}].}
\end{figure}

\subsubsection{Ni-Mn-Sn alloys}
The EB was also reported by Li {\it et al}. in an another FM shape memory alloy with composition Ni$_{50}$Mn$_{36}$Sn$_{14}$ \cite{li-EB-APL}. Analogous to that observed in Ni-Mn based alloys, Ni-Mn-Sn alloy showed a strong temperature dependence of EB effect that decreases sharply with increasing temperature and vanished well below the martensitic transition. Systematic change of EB effect has been reported in Mn-rich Ni-Mn-Sn alloys with composition Ni$_{50}$Mn$_{50-x}$Sn$_x$ as a result in variation of $x$ \cite{khan-EB-JAP}. The $H_E$ was found to decrease considerably with increasing $x$ from 14 to 16. This observation was suggested due to coexistence of AFM and FM coupling in the system. They further noted a double-loop structure while cooling the sample in ZFC condition and it  was attributed to a striped domain-type structure formed by FM regions of the system. 

The EB effect associated with the double-loop structure was also noticed in more Mn-rich alloys, Ni$_{50}$Mn$_{50-x}$Sn$_x$ with $x$ = 9 and 11 \cite{chat-EB-JPC}. It was observed that EB effect was considerably larger for $x$ = 11 than $x$ = 9. Strong $H_{cool}$ dependence of $H_E$ was observed which exhibited a sharp rise with increasing $H_{cool}$ initially, decreases sharply showing a peak in $H_{cool}$ dependence of $H_E$ and then it shows a slow decreasing trend with further increase in $H_{cool}$. Strong frequency dependence of ac susceptibility results clearly demonstrated SG-like behaviour close to EB blocking temperature, above which EB effect was  vanished. The EB effect involved with  RSG state at low temperature was reported in Ni$_{50}$Mn$_{34}$Sn$_{16}$ where RSG state was confirmed based on frequency dependence of ac susceptibility results \cite{chat-EB-PRB}. The ac susceptibility study indicated an onset of SG freezing near step-like anomaly with a clear frequency shift and RSG was envisaged in terms of coexisting FM and glassy magnetic phases at low temperature at least in field-cooled state. An example of the shift in magnetic hysteresis loop is demonstrated in figure 6(a) where the sample was cooled in different $H_{cool}$. Since a step-like feature was observed in temperature dependence of ZFC magnetization around 80 K, the sample was field cooled from 300 K and 80 K. Field cooling from 300 and 80 K practically did not show any significant difference which is highlighted in figure 6(b). When sample was cooled  from 40 K which was below the step-like feature (spin freezing temperature) in temperature dependence of ZFC magnetization, EB effect was not observed. Thus, the EB effect was proposed due to pinning effect at the FM and SG-like interface, unlike rest of the reports found among Ni-Mn based magnetic shape memory alloys where EB effect has been suggested at the FM/AFM interfaces. The SG-like features in the frequency dependence of ac susceptibility was suggested due to spin-frustration attributed to competing FM and AFM interactions at the interfaces. 

\subsection{Open issues}
The EB effects have been investigated in few binary alloys, Laves phase compounds and Heusler alloys where different aspects of magnetism were focused from EB effect. Evidence of EB effect close to zero magnetization in temperature dependence of ZFC magnetization and change of sign of $H_E$ associated with the change of sign of magnetization observed in few rare earth intermetallic compounds are fascinating  \cite{EB-thamiz-TOM,EB-thamiz-EPL}. Because it can probe coexisting magnetic phases close to significant temperature correlated to the magnetism. The reported experimental results in binary alloys arise some fundamental issues. Kouvel and his group clearly demonstrated evidence of EB phenomenology in diluted binary alloys such as Cu-Mn and Ag-Mn alloys \cite{kou,kou2} which has been recognised as canonical SG systems \cite{mydosh}. This opened up a fundamental issue to reinvestigate magnetic ground state of the canonical SG system or to reinvestigate mechanism of the EB phenomenology in classical SG  systems. Thus, EB mechanism in variety of  classical SG alloys and oxides needs to be explored extensively.   

It is noteworthy that evidences of EB effect in Ni-Mn based shape memory alloys are involved with Mn-rich compositions. It has been argued that short-range AFM correlation actually develops and/or becomes stronger in the martensitic phase of Mn-rich Ni-Mn-based shape-memory alloys \cite{enko-AFM,kren-AFM,brown-AFM}. However, signature of EB phenomenology in Mn-rich alloys further confirms existence of AFM interactions in the martensitic phase in addition to FM interaction. Until now, few literatures are reported mainly focused on Ni-Mn based Heusler alloys which have been reviewed herein. Still plenty of scopes of exploring EB effect in the magnetic shape memory alloys remain open for understanding the underline mechanism in magnetic shape memory alloys. 

\section{Exchange bias effect in oxide materials}

\subsection{Exchange bias in spontaneously phase separated oxides}
\subsubsection{Mixed-valent manganites having perovskite structure}
\paragraph{A. Charge ordered manganites}

\begin{figure}[t]
\centering
\includegraphics[width = 9 cm]{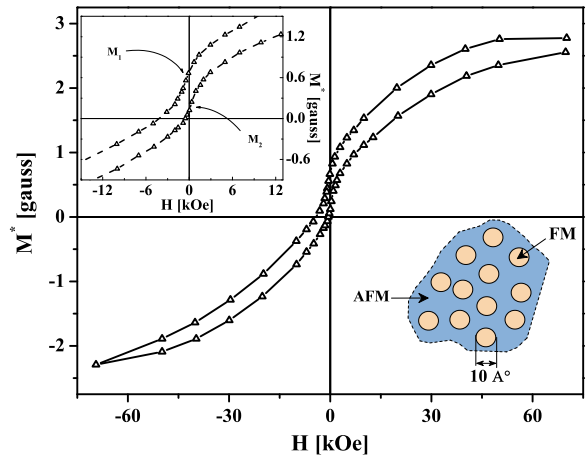}
\caption {Resulting magnetization $M^*$ vs $H$ loop at 5 K after subtracting the contribution of AFM matrix and cooling in 70 kOe for charge ordered manganite,  Pr$_{1/3}$Ca$_{2/3}$MnO$_3$. Inset highlights the low-field region, showing the shifts in the horizontal and vertical axes [reprinted with permission from \cite{nieb}]. The cartoon  in the lower inset is added to original figure which exhibits the proposed spontaneous phase separation scenario. Short range FM clusters are embedded in the AFM matrix. Average size of the FM clusters were estimated to be $\sim$ 10 \AA.}
\end{figure}

First evidence of EB effect in mixed valent manganites having perovskite structure  was reported in Pr$_{1/3}$Ca$_{2/3}$MnO$_3$ by Niebieskikwiat and Salamon \cite{nieb}. This signature in a spontaneously phase separated system created a renewed attention for investigating EB effect in structurally single-phase   compounds. When the sample was field cooled down to 5 K, a shift in magnetic hysteresis loop was noticed after subtracting the linear AFM component from experimentally obtained hysteresis loop. An example of shifted loop in the field (horizontal) and magnetization (vertical) axes are shown in figure 7. Training effect was noticed which was satisfactorily analysed by equation (1) for $\lambda \geq$ 2 and equation (2). In order to correlate the relation between horizontal and vertical shifts they proposed a simplified exchange interaction model where single domain FM clusters or droplets were embedded in an AFM matrix, analogous to that considered for single domain FM nanoparticles embedded in a non-FM matrix. Since it behaves like a single domain FM particles without interparticle interaction, the magnetization reversal takes place having forward switching frequency ($\nu_{+}$) and backward switching frequency ($\nu_{-}$). The number of particles switching forward (backward) is given by $\nu_{+} \tau$ ($\nu_{-} \tau$), where $\tau \sim 10^2 - 10^3$ s is typical measurement time. Thus, the change of magnetization associated with switching process is 
\begin{equation}
M_E/M_S = (\nu_{+} - \nu_{-})\tau
\end{equation} 
where $\nu_{\pm} = \nu_0 \exp[U_{\pm}/k_{B}T]$ with $U_{\pm} \approx KV \pm \mu H_E$. Here, $K$ is the anisotropy constant per unit volume, $V$ the volume of FM cluster, $\mu$ the magnetic moment of FM cluster, $k_{B}$ the Boltzmann constant and $\nu_0$ the switching attempt frequency typically found to be $\sim 10^9$ s. Replacing above parameters in equation (4) the expression is given by
\begin{equation}
M_E/M_S = -2\nu_0\tau \exp(KV/k_{B}T)\sinh(\mu H_E/k_{B}T). 
\end{equation} 
If $M_E, H_E, M_S$ and $\mu$ are known, $K$ can be determined from the above expression. For $\mu H_E < k_{B}T$ equation (5) can be found in simplified form as 
\begin{equation}
M_E/M_S \propto -H_E. 
\end{equation} 
For Pr$_{1/3}$Ca$_{2/3}$MnO$_3$ the ratio, $\mu H_E/k_{B}T$ was found to be lower than 0.9 which indicated that short range FM clusters embedded in the AFM matrix had a single domain character in this charge ordered compound. Niebieskikwiat and Salamon further proposed a simplified formula to characterize $H_{cool}$ dependence of the shifts in terms of $H_E$ or $M_E$ as 
\begin{equation}
-H_E \propto M_E/M_S \propto J_i\left[\frac{J_i\mu_0}{(g\mu_B)^2} L\left(\frac{\mu H_{cool}}{k_{B}T_f}\right) + H_{cool}\right].
\end{equation}
$J_i$ is the interface exchange constant, $g \approx$ 2 the gyromagnetic ratio, $\mu_B$ the Bohr magnetron. $\mu_0$ is the magnetic moment of core Mn spin where $\mu$ is moment of the FM clusters having $\mu$ = $N_v\mu_0$ and $N_v$ is number of spins within the FM clusters. The fit of $H_{cool}$ dependence of the shifts with equation (7) provides the value of $J_i$ and $N_v$. Average size of short range FM clusters are obtained from $N_v$ which was found to be $\sim$ 10 \AA ~for Pr$_{1/3}$Ca$_{2/3}$MnO$_3$ \cite{nieb} in accordance with the small angle neutron scattering results  \cite{gran-ND,henn-ND1,henn-ND2,henn-ND3}. Thus, systematic investigation of EB effect can provide quantitative estimate in accordance with the results obtained from neutron scattering experiment. It is important to point out that the experimental results and discussions of EB effect in  Pr$_{1/3}$Ca$_{2/3}$MnO$_3$ was reported based on the vertical shifts at field, $H$ = 0. This shift is often source of errors for calculating $M_E$ in the EB system. For example, Nogues and his coworkers demonstrated that the vertical shift at different $H_{max}$ showed a decreasing trend with increasing $H_{max}$ and it vanished at $H_{max}$ = 70 kOe  for MnO/Mn$_3$O$_4$ nanostructures \cite{noguesJACS}. Whereas a considerable stabilized value of the horizontal shift was observed at $H_{max}$ = 70 kOe, confirming the EB effect. To avoid ambiguity ascribed to minor loop effect it is reasonable to estimate the vertical shift at $M_S$ \cite{nogues-ver-shift}. In case of structurally single-phase alloys and compounds magnetization does not often saturate even at large $H_{max}$. In such a case plot of the vertical shift at different $H_{max}$ settles  the issue where stabilized value of the vertical shift at large $H_{max}$ confirms the EB effect.  

The signature of EB effect was noticed in an another charge ordered manganite, Nd$_{0.5}$Sr$_{0.5}$MnO$_3$ where EB effect was attributed to coupling between FM and AFM clusters \cite{prok-EB-JMMM}. The coexistence of Griffiths phase and EB effect has been observed in nanocrystalline La$_{1-x}$Ca$_{x}$MnO$_3$ ($x$ = 0.50, 0.67 and 0.75) where charge ordered state in the bulk counterpart were suppressed by decreasing grain size \cite{zhou-EB-JAP}. Moreover, field dependent magnetization revealed that short-range AFM region was still present in those nanoparticles which coupled to FM phase developed in nanocrystalline compound, leading to the EB phenomenology. 

\paragraph{B. Exchange bias ascribed to the rare-earth moment in  Eu$_{0.5}$R$_{0.5}$MnO$_3$ ($R$ = Pr and Sm)}

\begin{figure}[t]
\centering
\includegraphics[width = 8 cm]{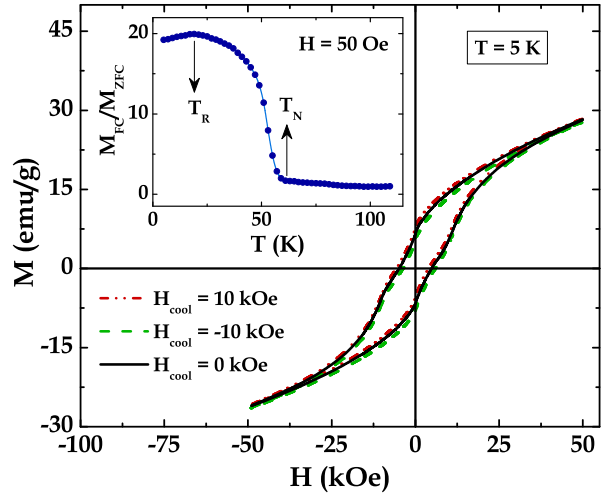}
\caption {Positive and negative shifts of magnetic hysteresis loop measured at 5 K for polycrystalline  Eu$_{0.5}$Pr$_{0.5}$MnO$_3$ after cooling the sample in 10 and -10 kOe field. Shift is absent for zero-field cooling. Inset shows plot of $M_{FC}/M_{ZFC}$ vs $T$ where disordered AFM transition ($T_{N}$) and transition ascribed to rare-earth spin ($T_{\rm R}$) are highlighted by the arrows \cite{arindam}.}
\end{figure}

Recently, a strong evidences of EB were observed in new series of bulk polycrystalline manganites with composition, Eu$_{0.5}$Pr$_{0.5}$MnO$_3$ \cite{arindam} and Eu$_{0.5}$Sm$_{0.5}$MnO$_3$ \cite{arindam1}. An example of the negative and positive shifts for 10 and -10 kOe cooling field is shown in figure 8 for Eu$_{0.5}$Pr$_{0.5}$MnO$_3$  which is typical manifestation of the EB effect. The shift was absent for zero-field cooling. A large $H_E$ of $\sim$ 1 kOe was observed at 5 K for $H_{cool}$ = 5 kOe. Interestingly, training effect was absent which is demonstrated in figure 9 at 5 K. Inset of figure further highlights that hysteresis loop exactly repeats even after third field cycling. The results were suggested due to stable interface moment involved with the EB phenomenon. It was noted that $H_E$ vanished above $T_{R}$ corresponding to ordering of rare-earth moment where EB effect was suggested due to pinning effect at the interface between disordered AFM and FM components. The signature of FM component was observed below $T_{R}$ in ZFC magnetization ($M_{ZFC}$). It is to be noted that signature of $T_{R}$ was observed by broad maximum in the plot of $M_{FC}/M_{ZFC}$ vs. $T$ which shown in the inset of figure 8. $M_{ZFC}$ and $M_{FC}$ are the ZFC and FC magnetization. Since anisotropy of rare-earth moment is typically larger than Mn-moment, $H_E$ reasonably vanished close to $T_R$, analogous to that observed in Co/CoO nanostructures where $T_N$ corresponding to highly anisotropic AFM CoO is less than $T_C$ corresponding to less anisotropic FM Co. The EB effect in Eu$_{0.5}$Sm$_{0.5}$MnO$_3$ also displays similar results where FM-like transition at $T_R$ in Eu$_{0.5}$Pr$_{0.5}$MnO$_3$ exhibits FIM behaviour in Eu$_{0.5}$Sm$_{0.5}$MnO$_3$ \cite{arindam1}.

\paragraph{C. Other manganites}
The EB effect was recently reported in single crystal of Gd$_{2/3}$Ca$_{1/3}$MnO$_3$. This  was attributed to spontaneously phase separated inhomogeneous ferrimagnetism \cite{haber-EB-JMMM}. The EB effect was strongly dependent on direction of  external magnetic field which changed its sign at the compensation temperature around 16 K. The EB effect was observed in a spontaneous lamellar FM/AFM phase separated manganite, Y$_{0.2}$Ca$_{0.8}$MnO$_3$  \cite{qian-EB-APL}. The $H_{cool}$ dependence of $H_E$ showed that $H_E$ was decreased to 37 \% of the value for increasing $H_{cool}$ from 10 kOe to 60 kOe. The $H_E$ was also found to be proportional to inverse of $M_S$, suggesting that $H_E$ was inversely proportional to the size of FM domain. The EB effect associated with the CG-like and SG-like states has been reported in  $L_{0.5}$Sr$_{0.5}$MnO$_3$ ($L$ = Y, Y$_{0.5}$Sm$_{0.5}$ and Y$_{0.5}$La$_{0.5}$) \cite{BKC-EB-PRB}. Training effect was observed and analysed by equation (1) for $\lambda \geq$ 2 and equation (2). Cooling field dependence of EB effect showed that $H_E$ was found to increase sharply and then it decreased due to growth of FM cluster. The EB  effect was suggested at the FM and SG-like interfaces. 

\begin{figure}[t]
\centering
\includegraphics[width = 8 cm]{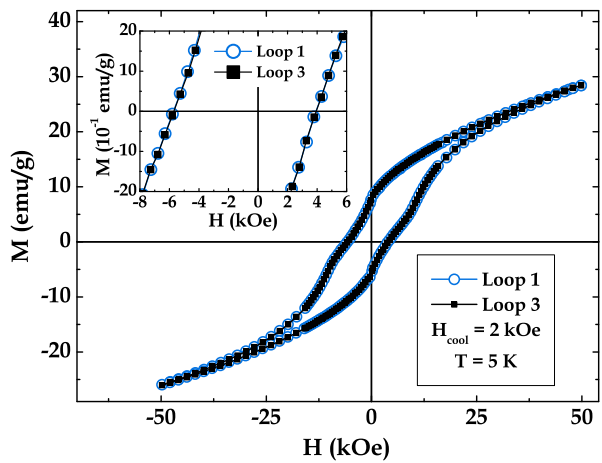}
\caption {Magnetic hysteresis loop measured at 5 K for polycrystalline  Eu$_{0.5}$Pr$_{0.5}$MnO$_3$ after first (Loop 1) and third (Loop 3) field cycling exhibiting absence of training effect. Inset highlights central part of hysteresis around the origin  \cite{arindam}.}
\end{figure}

The effect of Mn-site substitution and EB effect have been investigated in La$_{1-y}$Mn$_{1-x}$Fe$_{x}$O$_3$ \cite{patra-EPJB07,de-JPD08,thakur-JPCM08} and Bi$_{0.4}$Ca$_{0.6}$Mn$_{1-x}$Ti$_x$O$_3$ \cite{zhu-JPD09}.  First report of EB effect as a result of Mn-site substitution was found in LaMn$_{0.7}$Fe$_{0.3}$O$_3$ \cite{patra-EPJB07}. It was noticed that EB effect was observed only at 30 \% of Fe substitution which was absent for other compositions with different percentage of Fe content. The model proposed in  CG-like state comprised of short range FM clusters embedded in a SG-like matrix was suggested at low temperature for this compound where EB was proposed due to pinning effect at the FM and SG interface. The $H_E$ was vanished close to spin-freezing temperature. The average size of FM cluster was $\approx$ 25 \AA ~ as estimated from the analysis of $H_{cool}$ dependence of $H_E$ at 5 K using equation (7). The $H_E$ was found to decrease up to 45 \% for an increase of $H_{cool}$ from 5 kOe to 12 kOe. Grain size effect on EB was investigated on LaMn$_{0.7}$Fe$_{0.3}$O$_3$ with average size 20, 90 and 300 nm \cite{thakur-JPCM08}. The $H_E$ decreased considerably with an increase of particle size and weak EB effect was still observed for the particle with 300 nm average size. It was notable to point out that finite size effect of $H_E$ as a function of $H_{cool}$ showed anomalous behaviour. The results are displayed in figure 10. The values of $H_E$ were considerably larger for the particles with 90 nm size than the particles having 20 nm size for $H_{cool} \leq$ 4 kOe. The values of $H_E$ for the particles having 20 nm size overshoot the value of $H_E$ of the particles with 90 nm size for $H_{cool} >$ 4 KOe. The results clearly demonstrate that grain size effect of $H_E$ strongly depends on cooling field. It has also been noticed that larger $H_E$ was involved with larger $H_C$ which is displayed in the inset of figure 10. Inset of the figure further showed that initially increase of $H_E$ is associated with increase of $H_C$, The slope of increase of $H_C$ is considerably decreased while $H_E$ decreases with $H_{cool}$ in the high-field region of the plot.  
 With increasing grain size average size of FM clusters was proposed to increase, resulting in considerable decrease of $H_E$. From the fits of $H_{cool}$ dependence of EB effect using equation (7) average sizes of the short range FM clusters were $\approx$ 10 and $\approx$ 30 \AA ~for the samples with grain size 20 and 90 nm, respectively. The $H_{cool}$ dependences of $H_E$ and fits of the experimental data using equation (7) are shown in figure 10. Signature of EB effect was also noticed in La$_{0.87}$Mn$_{0.7}$Fe$_{0.3}$O$_3$ \cite{de-JPD08}. The effect was found to be weaker for La-deficient, La$_{0.87}$Mn$_{0.7}$Fe$_{0.3}$O$_3$ than La-stoichiometric, LaMn$_{0.7}$Fe$_{0.3}$O$_3$ where weaker $H_E$ was proposed to be associated with the larger size ($\approx$ 63 \AA) of short range FM clusters. Thus results of EB phenomenon provide a significant information that La-deficiency in these compounds leads to the increase of FM clusters. The EB has also been investigated in charge ordered compound, Bi$_{0.4}$Ca0$_{0.6}$Mn$_{1-x}$Ti$_{x}$O$_3$ (0 $\leq x \leq$ 0.2) \cite{zhu-JPD09}. The EB parameters were strongly depending on Ti substitution. In particular, $H_E$ showed a nonmonotonic variation with a maximum at Bi$_{0.4}$Ca$_{0.6}$Mn$_{0.92}$Ti$_{0.08}$O$_3$. The EB has been interpreted in terms of density of FM nanodomains superposed on the AFM matrix. The results suggested that it was possible to tune EB through Mn-site substitution in charge ordered manganites by controlling density of FM nanodomains. 

\begin{figure}[t]
\centering
\includegraphics[width = 7.5 cm]{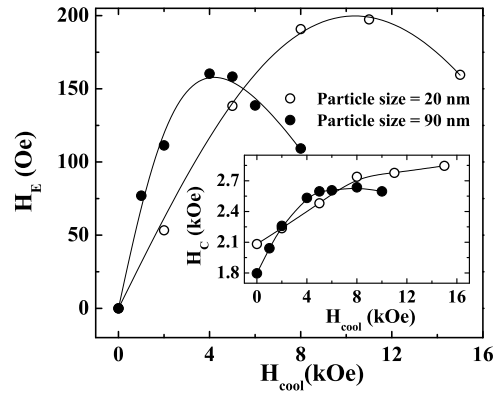}
\caption {The $H_{cool}$ dependence of $H_E$ for LaMn$_{0.7}$Fe$_{0.3}$O$_3$ having 20 and 90 nm average size, respectively. The solid lines show the fits as described in equation (7). The inset of the figure exhibits coercivity ($H_C$) against $H_{cool}$ for both the particles. The solid lines are the guides for eye [reprinted from \cite{thakur-JPCM08}].}
\end{figure}

\subsubsection{Mixed-valent cobaltites having perovskite structure}
\paragraph{A. La$_{1-x}$Sr$_{x}$CoO$_3$}
The mixed-valent cobaltites with perovskite structure experience a delicate interplay between charge, spin state, transport, magnetic and structural degrees of freedom, exhibiting a complex phase separation scenario \cite{wu-PS-PRB,stau-PS-PRB}. First evidence of EB effect was attributed to the spontaneous phase separation as reported in cobaltites, La$_{1-x}$Sr$_{x}$CoO$_3$ by Tang {\it et al}. \cite{tang-EB-PRB,tang-EB-JAP}. Training effect was investigated and was satisfactorily analysed by using equation (1) for $\lambda \geq$ 2 and equation (2). Maximum EB effect was found in La$_{0.88}$Sr$_{0.12}$CoO$_3$ where $H_E$ was decreased to 40 \% for an increase of $H_{cool}$ from 20 kOe to 50 kOe \cite{tang-EB-JAP}. In order to confirm EB effect attributed to grain interior spontaneous phase separation, investigation was performed in a single crystal with composition, La$_{0.82}$Sr$_{0.18}$CoO$_3$ where EB effect was ascribed to pinning effect at the FM and SG interface \cite{huang-EB-JPCM}. The analysis of $H_{cool}$ dependence of $H_E$ using equation (7) provided coupling constant $\approx$ -0.68 meV at the interface between FM and SG phases and average size of FM clusters $\approx$ 0.9 nm. The $H_{cool}$ and $H_{max}$ dependences of EB effect were further investigated in La$_{0.82}$Sr$_{0.18}$CoO$_3$ \cite{yuaLSCO}. The $M_E$ was found to vanish whereas $H_E$ was $\sim$ 450 Oe at 3 K for $\left|H_{max}\right| \geq$ 40 kOe. They further investigated  $H_{cool}$ dependence of training effect which could be analysed by equation (1) and equation (2). The results indicated suppression of SG region with increasing $H_{cool}$ that played a significant role in decrease of $H_E$.   

\begin{figure}[t]
\centering
\includegraphics[width = 7.5 cm]{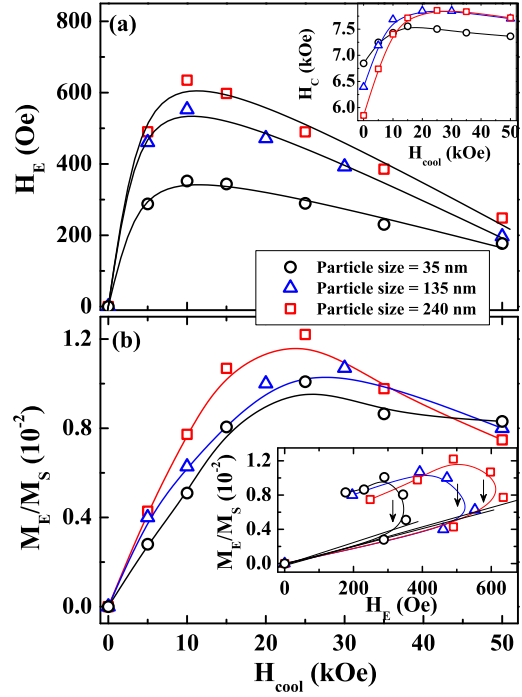}
\caption {The $H_{cool}$ dependence of (a) $H_E$ and (b) $M_E/M_S$, at 5 K with particle size 35, 135 and 240 nm for La$_{0.88}$Sr$_{0.12}$CoO$_3$. The $H_{cool}$ dependence of coercivity ($H_C$) are shown in the inset of (a). Plots of $M_E/M_S$ against $H_E$ are shown in the inset of (b) [reprinted from \cite{patra-EB-JPCM10}].}
\end{figure}

\begin{figure}[t]
\centering
\includegraphics[width = 7.5 cm]{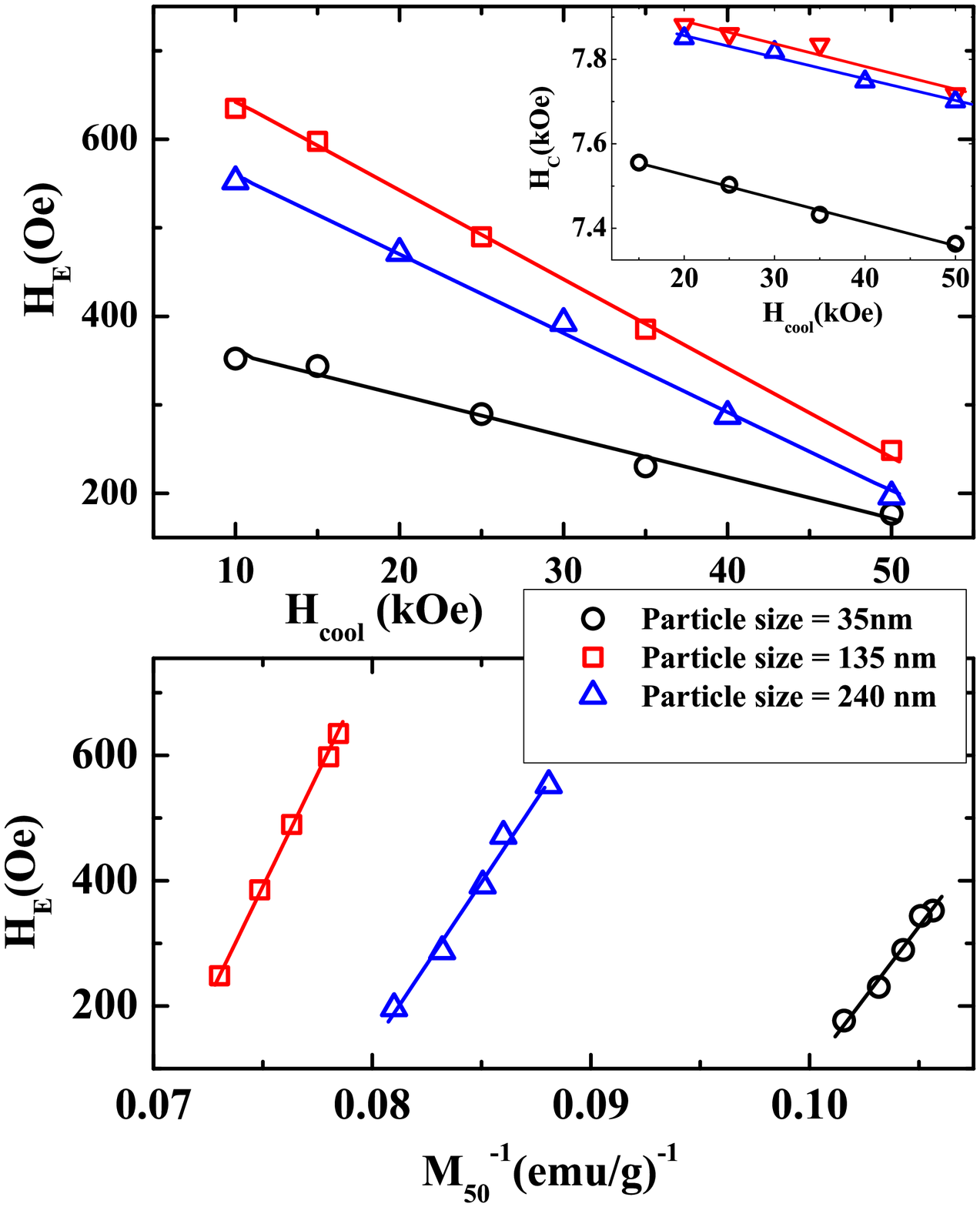}
\caption {(a) Plots of $H_E$ against $H_{cool} \geq$ 10 kOe at 5 K for La$_{0.88}$Sr$_{0.12}$CoO$_3$ with particle size 35, 135 and 240 nm. Inset shows plot of coercivity ($H_C$) against $H_{cool}$. (b) Plot of $H_E$ as a function of  inverse of average magnetization at 50 kOe ($M_{50}^{-1}$). Solid straight lines indicate the linear fits  [reprinted from \cite{patra-EB-JPCM10}].}
\end{figure}

Grain size effect on magnetic CG properties of La$_{0.88}$Sr$_{0.12}$CoO$_3$ was investigated for the samples with 35, 135 and 240 nm grain sizes where systematic observation of EB effect was exploited to probe microscopic insight of the grain interior spontaneous phase separation scenario \cite{patra-EB-JPCM10}. The bulk compound is a CG compound composed of short range FM clusters embedded in the SG matrix at low temperature \cite{wu-PS-PRB}. The short range FM clusters were still retained in the nanocrystalline form with average size 35 nm associated with the SG component which was  confirmed by  considerable EB effect. The EB effect manifested by the shift of hysteresis loop was observed due to field cooling where the effect was found to be weakened monotonously  with decreasing grain size. The decrease in fraction of FM component was found to be correlated with the weakening of EB effect with decreasing grain size. The $H_{cool}$ dependence of $H_E$ and $M_E$ scaled by $M_S$ ($M_E/M_S$) were investigated (figure 11). Inset of figure 11(a) exhibits the plots of $H_C$ with $H_{cool}$ where initial sharp rise of $H_E$ are associated with increase of $H_C$. Eventually $H_C$ shows a decreasing trend with increasing $H_{cool} \geq$ 10 kOe, analogous to that observed in the $H_E$ vs $H_{cool}$ plot. The plot of $M_E/M_S$ shows similar $H_{cool}$ dependence up to 10 kOe which is demonstrated in the inset of figure 11(b) by linear plot of $M_E/M_S$ against $H_E$. For $H_{cool} \geq$ 10 kOe plot deviates from the linearity. According to equation (6) linear plot for $\mu H_E < k_B T$ indicates single domain structure of the FM clusters. It was suggested that growth of FM clusters and $\mu$ were increased with increasing $H_{cool}$ where the relation, $\mu H_E < k_B T$ did not hold for $H_{cool} \geq$ 10 kOe.  Furthermore, $H_E$ [figure 12(a)] and $H_C$ [inset of figure 12(a)] were found to decrease linearly with $H_{cool}$ above 10 kOe. It is notable that $H_E$ decreases to 60\% for sizes $\sim$ 135 nm and $\sim$ 240 nm whereas it decreases to 50\% for 35 nm with the increase of $H_{cool}$ from 10 to 50 kOe. The decrease of $H_E$ was suggested to be associated with growth of FM clusters in accordance with the model proposed by Meiklejohn and Bean \cite{meik3}
\begin{equation}
H_E = J_{ex}/(M_{FM} \times t_{FM}).
\end{equation} 
$J_{ex}$ is the exchange constant across the FM/AFM interface per unit area, $M_{FM}$ the magnetization and $t_{FM}$ the thickness of the FM layer. The increase of $M_{FM}$ associated with increase of $t_{FM}$ in denominator of the above expression decreases the magnitude of $H_E$. It was further suggested that increase of M$_{50}$ (magnetization at 50 kOe) with increasing $H_{cool}$ indicated the rise of average size of FM clusters due to  increase of $H_{cool}$ for 
$H_{cool} >$ 10 kOe. The values of $H_E$ against 1/$M_{50}$ are plotted in figure 12(b) for all the particles. It was evident from the plots that the values of $H_E$ were inversely  proportional to $M_{50}$, i.e. average size of FM clusters. The slope defined as  $dH_{E}/d(M_{50}^{-1})$ was found to increase with increasing grain size [see figure 12(b)], indicating that EB effect for particles having largest average size is more sensitive to inverse of $M_S$ or $H_{cool}$.

\paragraph{B. Nd$_{1-x}$Sr$_{x}$CoO$_3$}

Similar to that observed in La$_{1-x}$Sr$_{x}$CoO$_3$, the series of cobaltites, Nd$_{1-x}$Sr$_{x}$CoO$_3$ also displays spontaneous phase separation scenario  \cite{stau-PS-PRB}. For low doping range (0 $< x <$ 0.18) SG or CG state
has been proposed with resistivity showing a semiconducting temperature dependence. With further increase in hole doping the short range FM clusters begin to coalesce above 
a percolation threshold ($x >$ 0.18) to attain magnetic long range ordering and start to show metallic conductivity in the ordered state. The coexistence of FIM and FM ordering was  reported for 0.20 $\leq x \leq$ 0.60. The spontaneous phase separation scenario and EB effect were recently investigated in 
Nd$_{1-x}$Sr$_{x}$CoO$_3$ \cite{patraJPCM1,patraSSC}. Below the percolation threshold the EB effect was attributed to pinning effect at the interface between  FM and SG phases \cite{patraSSC}. The $H_E$ and $M_E$ showed training effect which was in accordance with equation (1) for $\lambda \geq$ 2 and equation (2). The plot of $M_E/M_S$ against $H_E$ was linear for $H_{cool}$ in the range 0 $\leq H_{cool} \leq$ 50 kOe which indicated single domain structure of the FM clusters below the percolation limit. The EB effect was also performed at $x$ = 0.20 and 0.40 where maximum EB effect was observed close to the percolation threshold at $x$ = 0.20 \cite{patraJPCM1}. This was attributed to FM and FIM interface where $H_E$ vanished close to FIM ordering temperature. It was noted that the linear plot of $M_E/M_S$ with  $H_E$ was also observed at $x$ = 0.20 close to the percolation threshold which confirmed single domain structure of the FM clusters. In fact, single domain structure of FM clusters was also observed even well above the percolation threshold at $x$ = 0.40. The analysis of $H_{cool}$ dependence of $M_E/M_S$ using equation (7) provided average size of the FM clusters $\sim$ 20 \AA ~and $\sim$ 40 \AA ~for $x$ = 0.20 and 0.40, respectively.   

\paragraph{C. Pr$_{1-x}$Sr$_{x}$CoO$_3$}

The EB effect was recently observed in Pr$_{1-x}$Sr$_{x}$CoO$_3$ close to the percolation threshold ($x$ = 0.25) at $x$ = 0.20 and 0.30 where stronger effect was noticed in $x$ = 0.30 \cite{patraJAP}. Training effect was observed which could be analysed by equation (1) for $\lambda \geq$ 2 and equation (2). The linear behaviour of $H_E$ with $M_{E}/M_{S}$ in agreement with equation (6) indicated single domain structure of the FM clusters where EB effect was attributed to pinning effect at the FM/SG interface. The considerable decrease of $H_E$ ($\sim$ 66\%) was observed for an increase in $H_{cool}$ from 7.5 to 50 kOe. The decrease of $H_E$  above $H_{cool}$ = 7.5 kOe was correlated with the growth of FM clusters in accordance with equation (8). 

\subsubsection{Other oxides with perovskite structure}
\paragraph{A. Ferrites}
Signature of EB effect in multiferroic BiFeO$_3$ film \cite{all-APL09,leb-PRB10} opens up a fundamental question whether structurally single-phase BiFeO$_3$ is a magnetically inhomogeneous material. The EB effect has also been observed in bulk, Bi$_{1-x}$La$_{x}$FeO$_3$ with $x$ = 0.05, 0.10, 0.15 and 0.20 over wide temperature range 2 - 300 K \cite{das-JAP07}. The EB was found in polycrystalline,  Pr$_{1-x}$Sr$_{x}$Fe$_{0.8}$Ni$_{0.2}$O$_{3-\delta}$ with $x$ = 0.1, 0.2 and 0.3 where a frustrated magnetic order ascribed to competing FM and AFM interactions were proposed to be correlated with the EB effect \cite{larr-JAP08}. A significantly large EB effect ($H_E \approx$ 1.44 kOe at 10 K) has been reported in oxygen deficient, SrFeO$_{2.75}$ which was suggested to be involved with Fe(2) moment N\'{e}el temperature ($\sim$ 230 K), below which the EB appeared due to FC process \cite{will-PRB09}.

\paragraph{B. Titanates}
The EB has been reported in structurally single-phase compound, BaTiO$_3$ as a result of Ba-site \cite{luo-APL08} doping and Ti-site \cite{luo1} substitution. The ferromagnetism was proposed in ferroelectric compound with composition  Ba$_{0.5}$Sr$_{0.5}$Ti$_{0.97}$Co$_{0.03}$O$_{3}$ where ferromagnetism was attributed to Co substitution \cite{luo-APL08}. However, signature of EB effect confirmed the existence of magnetic inhomogeneity associated with ferromagnetism. Training effect was noticed in the above compound. Similar EB and training effects were also reported in BaTi$_{0.98}$Co$_{0.02}$O$_3$ film where magnetic inhomogeneity was suggested to interpret the EB effect \cite{luo1}.

\subsubsection{Zn$_{x}$Mn$_{3-x}$O$_4$}
The EB effect has been investigated in bulk specimens of the solid solution,  Zn$_{x}$Mn$_{3-x}$O$_4$ with $x$ = 0, 0.25, 0.5, 0.75 and 1 where structural phase-purity of the  polycrystalline samples was ascertained by neutron-diffraction measurements \cite{shoe-PRB09}. Samples with $x$ = 0.25, 0.5 and 0.75 exhibited shifted magnetic hysteresis loops at low temperature, characteristic of EB typically seen in magnetic composites. They proposed that unusual magnetic behaviour appeared due to nanoscale mixture of short range FIM and AFM regions. They further suggested that the results found in the solid solution proposed the insights into alloying of a FIM Mn$_3$O$_4$ with AFM ZnMn$_2$O$_4$ wherein distinct magnetic clusters could grow and percolate to produce a smooth transition between competing orders.

\subsection{Concluding remarks}
The EB effects attributed to the spontaneous phase separation are mainly observed in Manganites, Cobaltites, Ferrites and Titanates having different perovskite structures. These observations provide plenty of qualitative and quantitative information about nanoscale phase separation in the grain interior using macroscopic experimental result such as magnetization which are in accordance with the results obtained from the microscopic experiments viz., neutron scattering or nuclear magnetic resonance, etc. In few compounds FM clusters or droplets embedded in non-FM matrix were proposed where rough estimates of the FM size was determined from the simplified exchange interaction model. However, these values of FM size are in same order of magnitude obtained from the neutron scattering experiments. In few of the cases of spontaneously phase separated systems contrast behaviour of grain size dependence of $H_E$ is noticed which is significant to probe qualitative depiction of the nanoscale magnetic phase separation scenario. Specially, grain interior magnetic phase separation scenario of CG compounds composed of short range FM clusters embedded in the SG-like matrix can be explored by changing the grain size. In case of structurally single phase compound and core-shell structures having different structural phases non-monotonous grain size dependence of EB is often noticed which is not clearly understood. The $H_{cool}$ dependence of $H_E$ for different grain size displayed in figure 10 is interesting in this context which clearly exhibits that grain size effect of $H_E$ strongly depends on $H_{cool}$. Therefore, $H_{cool}$ dependence of the EB  effect for different grain sizes needs to be investigated extensively for understanding non-monotonous behaviour of grain size dependence of the EB effect. 

Most of the results involved with SG or SG-like magnetic phase clearly exhibit a considerable decrease of $H_E$ (even up to $\sim$ 66 \%) with increasing $H_{cool}$ in the high-field regime. This decrease has been suggested usually due to enhancement of FM size. This arises an open issue whether proportion of FM phase to SG or SG-like phase is increased by increase of $H_{cool}$? If it is so, then why decrease of $H_E$ is mainly limited to the compound associated with SG phase? Theoretical interpretation this result needs to explore extensively. 

\section{Exchange bias in magnetic core-shell nanostructures}

\subsection{Introduction}
In case of fine particle system surface to volume ratio becomes significantly large compared to bulk counterpart. In such a case surface effect dominates over the core part quite often, leading to variety of magnetism \cite{kodama-rev}. Because of enhanced surface effect having improved magnetic properties it has been recognized as an important class of functional materials \cite{liu-rev}. The  strong surface anisotropy typically gives rise to  EB effects which have been extensively reviewed on core-shell structures by Iglesias {\it et al}. where the experimental results were interpreted by existing theories and models \cite{nogues1,nogues2,iglesias,monte-ig1,monte-ig2,iglesias1,iglesias2}. 
Current experimental results indicate that evidence of EB is not limited only to the core-shell structures. These have been reported in various combinations of metal and metal oxides. Few interesting examples reported very recently are given below. The EB was reported in Fe/Fe oxide nanogranular systems displaying glassy dynamics associated with the EB phenomenology \cite{Fe-FeO1,Fe-FeO2,Fe-FeO3}. The Fe/Fe oxide system was composed  Fe particles (average size $\sim$ 6 nm) dispersed in a structurally and magnetically  disordered iron oxide matrix (mixture of Fe$_3$O$_4$ and $\gamma$-Fe$_2$O$_3$ having average size $\sim$ 2 nm). Interestingly, $H_E$ was found to increase with increasing time  spent at low temperature after applying cooling field. The authors interpreted that oxide phase evolves towards a lower energy configuration during time delay, resulting in a stronger interface exchange coupling with the Fe particle moments. This was confirmed by Monte Carlo simulation assuming a random shell anisotropy \cite{Fe-FeO3}. Finite size effect has been investigated in naturally oxidized Fe nanoparticles \cite{Fe-FeO4}. It was demonstrated that $H_E$ decreased with decreasing average particle size and vanished below a critical size. The oxide phase was $\gamma$-Fe$_2$O$_3$ for size below $\sim$ 5 nm and an additional Fe$_3$O$_4$ phase appeared for larger particle size. The shell driven magnetic stability was proposed in typical Co/CoO having core-shell structure \cite{Co-CoO1}. The superparamagnetic blocking temperature, $H_C$ and $H_E$ were found to increase with increasing coverage densities. The authors pointed out that the shells of isolated core-shell nanoparticles was strongly degraded which were rapidly recovered as nanoparticles came into physical contact. The controll of EB has been demonstrated in a Co-core/CoO-shell nanostructure \cite{Co-CoO2}. The authors pointed out that interplay between shell thickness and lattice strain induced net moment at the core-shell interface lead to controll of EB. The EB effect at the irregular interface between FM Co and AFM CoO has been reported in Co/CoO nanostructures where Co/CoO having typical core-shell did not require for the strong EB effect \cite{das}. The vertical shift was found to be uncorrelated with the horizontal shift in Ni/NiO nanostructures \cite{thakur1}. The horizontal shift was found to decrease with average particle size retaining the unchanged vertical shift. 
Mechanically ball milled Ni/NiO nanostructures displayed EB effect which was attributed to the exchange interaction between FM Ni and disordered NiO components \cite{Ni-NiO1,Ni-NiO2}. The content of Ni in Ni/NiO nanostructures and particle size were reported to be important parameters for EB effect \cite{Ni-NiO3}.  

The numerous reports on EB effects are found in the literatures based on metal/metal, oxide/oxide and metal/oxide having core/shell structure and different structural phases. Few recent examples of EB on metal/metal oxide nanostructures are briefly discussed in the subsection. However, detailed review on the EB phenomenon observed in variety of core-shell structures are beyond the scope of this review. Here, the motivation of this review is restricted on structurally single-phase alloys and compounds which are described in following two subsections: Magnetic core-shell structures in alloys and Magnetic core-shell structure in oxides.

\subsection{Core-shell magnetic structures in alloys}

The EB effect was investigated on ball-milled nanocrystalline, FeRh alloys having {\it fcc} structure \cite{hern-alloy-surf}. The shift in magnetic hysteresis loop was  observed after the field cooling which was described as manifestation of EB effect. The experiments were performed in three samples with average grain sizes $\sim$ 8.6, $\sim$ 13.2  and $\sim$ 17.3 nm. The $H_E$ was noticed to be $\sim$ 1610, $\sim$ 500 and $\sim$ 130 Oe, respectively indicating that $H_E$ increased considerably with decreasing average grain size, attributed to strongest surface effect for the smallest particle. Spin freezing at the grain boundaries behaving SG-like phase was suggested to be involved with the EB effect.

Mechanical milling of ordered FM SmCo$_5$ alloy exhibited a dramatic increase in $H_C$  associated with the loop shift at low temperature after FC process \cite{pele-alloy-surf}. This shift was suggested as a manifestation of EB effect. Time period of milling was found crucial for enhancement of $H_C$ as well as EB effect. Both of it were  found to increase  initially and then decreased with increase in milling time period, showing a peak around $\sim$ 2.5 h of milling time. The results further indicated that large $H_C$ was involved with large $H_E$. The high coercivities were attributed to formation of nanostructures composed of crystalline SmCo$_5$ core regions covered by the disordered magnetic grain boundaries.

The EB effect was observed in nanocrystalline, Fe$_{73.5}$Nb$_{4.5}$Cr$_5$B$_{16}$Cu$_1$ \cite{skor-alloy-surf}. The remarkable shift of hysteresis loops exhibiting EB effect was detected at liquid helium temperatures after the field cooling. The disordered magnetic phase at the grain boundaries was suggested to interpret the EB effect in nanocrystalline system. 

The strong signature of EB effect was observed in chemically synthesized CoNi alloy embedded in the silica matrix having 10 \% volume fraction \cite{das-alloy-surf}. As synthesized sample was heated at 500$^0$ C and 800$^0$ C in presence of atmospheric pressure of H$_2$/Ar gas mixture which resulted in tuning the average grain size \cite{thakur-alloy-surf}. Larger particles (average size $\sim$ 12.0 nm) heated at 800$^0$ C had a larger $M_S$ with smaller $H_C$ than that of smaller particles (average size $\sim$ 8.5 nm). Larger $H_C$ for the particles with 8.5 nm size was involved with the larger EB effect. The values of $H_E$ were 850 and 350 Oe for the particles with average size 8.5 and 12.0 nm, respectively. A typical demonstration of the shifts in hysteresis loops are displayed in figure 13 at 5 K when samples were cooled in $H_{cool}$ = 1 kOe. The central portions of the loops highlighted in the inset demonstrate  the distinct shifts of loops along the field axis. A strong correlation between dipole-dipole interaction and EB effect was found. The dipole-dipole interaction was controlled by tuning volume fraction of CoNi alloy retaining same particle size. It was noted that EB effect vanished while volume fraction of CoNi alloy was reduced to 0.01 \%. This volume fraction was close to superparamagnetic limit where $H_E$ was vanished.  

Recently, EB was observed in chemically synthesized PtNi nanoparticles of average size from 2.27 nm to 4.11 nm \cite{ho-alloy-surf}. Both the horizontal and vertical shifts were observed at 5 K in FC process. Unlike other reports on the EB effect attributed to the disordered magnetic state at grain boundary, EB effect in PtNi nanoparticles was suggested due to pinning effect at the interfaces between FM surface and AFM core phases. It was found that $H_C$ and $H_E$ were strongly dependent on particle size. The values of $H_E$ were 1200, 3000 and 1100 Oe for nanoparticles with average sizes 2.27,  3.09 and 4.11 nm, respectively.

\begin{figure}[t]
\centering
\includegraphics[width = 9 cm]{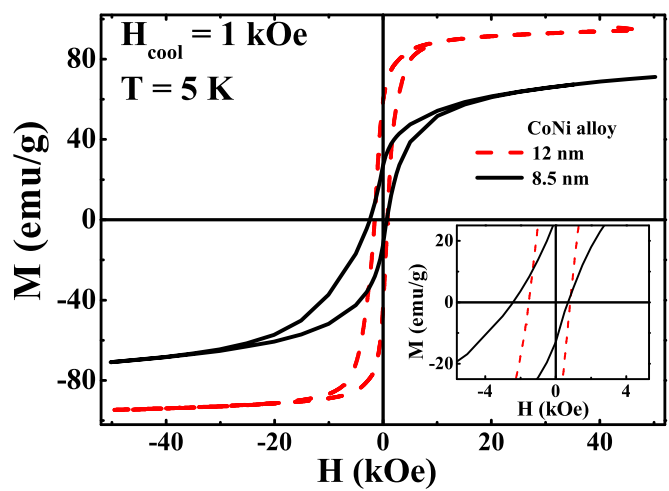}
\caption {Shifts of the magnetic hysteresis loops are displayed at 5 K after cooling the samples in $H_{cool}$ = 1 kOe in two CoNi alloys having 8.5 and 12.0 nm grain sizes. Inset highlights the central portions of the hysteresis loops, exhibiting the shifts along the field axis \cite{das-alloy-surf}.}
\end{figure}

\subsection{Magnetic core-shell structures in oxides}

\subsubsection{NiO}

The signature of large magnetic moments, coercivities and loop shifts up to 10 kOe was reported by Kodama {\it et al}. in an AFM NiO nanoparticles \cite{kodama-PRL97}. This anomalous behaviour involved with the finite size effect was interpreted initially by numerical modeling of spin configurations considering  multi-sublattice configurations \cite{kodama_theo}. Thus, the reduced coordination of surface spins was proposed for the fundamental change in magnetic order of the fine particle system. The relatively weak coupling between the sublattices allowing a variety of reversal paths for the spins upon cycling the applied field was suggested to interpret large $H_C$ and loop shifts for the fine particle of AFM NiO. Recently, Kodama and  his group extended their studies on EB effect in nanocrystalline NiO with sizes in between 5 nm to 55 nm where the loop shift ascribed to field cooling was suggested due to the EB effect \cite{kodama-SSC08}. The $H_E$ and FM magnetization arising from the uncompensated surface spins of otherwise AFM NiO showed inverse trends when presented it  as functions of particle size. An interfacial exchange energy $\sim$ 0.03 erg/cm$^2$, comparable to those found in NiO/Co and NiO/permalloy bilayers, was reported. 
Very recently, EB effect has also been investigated in the nanoparticles of NiO where different interpretations were proposed \cite{jago-JPCM09,sharma-nanotech10}. Performing experiments on the NiO nanoparticles of different sizes and bulk NiO, Jagodi$\check{c}$ {\it et al}. showed the coercivity enhancement and the shift of hysteresis loop after field cooling. This considered to be the key experimental manifestations of multisublattice ordering and the EB effect which were true nanoscale phenomena only present in the nanoparticles and absent in the bulk counterpart \cite{jago-JPCM09}. On the other hand, Sharma {\it et al}. proposed that the strong EB effect associated with the large loops shift $\sim$ 2.2 kOe and considerable enhancement of $H_C$ ($\sim$ 10.2 kOe) was attributed to the  interface coupling between AFM NiO core and disordered magnetic surface at the grain boundary \cite{sharma-nanotech10}. They further demonstrated linear dependence of the horizontal shift against the vertical shift and proposed the role of pinned spins on exchange fields from this linear dependence. 

\subsubsection{CuO}

The EB effect was investigated in AFM nanoparticles of CuO \cite{pun-PRB01,pun-JAP02,pun-SSC03,zheng-SSC04}. Punnoose {\it et al}. demonstrated that existence of uncompensated surface spins resulted in a weak FM component for the particles below $\sim$ 10 nm which caused an appearance of EB effect along with the increase of $H_C$ \cite{pun-PRB01,pun-JAP02}. Similar results have also been reported in AFM CuO based on the magnetization and muon spectroscopic  measurements which was  prepared by ball-milling a CuO single crystal \cite{zheng-SSC04}. 

\subsubsection{Iron oxides}

Nanoparticles of $\gamma-$Fe$_2$O$_3$ with a very high surface to volume ratio exhibited  strong exchange anisotropy as well as training effect \cite{mart-PRL98}. Temperature dependence of exchange anisotropy field showed that it vanished close to spin freezing temperature which has been interpreted in terms of  random-field model of exchange anisotropy. In the framework of this theory a surface SG layer about $\sim$ 0.6 nm thick was determined \cite{mart-PRL98}. The effect of surface spin disorder on magnetism of $\gamma-$Fe$_2$O$_3$  nanoparticle of different particle sizes  prepared from different chemical routes was investigated where surface spin disorder at the grain boundary was exchange coupled to the ordered core, giving rise to the EB effect \cite{shen-nanotech07,li-JMC07,maiti-JMMM09}. Li {\it et al}. demonstrated that $M_S$  increased whereas $H_C$ decreased with increasing average grain size \cite{li-JMC07}. It is noted that better crystallinity is indicated by large $M_S$ with small $H_C$ which is also involved with the small EB effect. Salazar-Alvarez {\it et al}. reported that EB effect was also dependent on shape of $\gamma-$Fe$_2$O$_3$ nanoparticles \cite{nogues-JACS-2008}. Experimental results and Monte Carlo simulations suggested that  different random surface anisotropies of the two morphologies combined with low magnetocrystalline anisotropy of $\gamma-$Fe$_2$O$_3$ were the origin of observed effect.  Recently, considerable EB effect has been reported in noninteracting $\gamma-$Fe$_2$O$_3$ hollow nanoparticles which was attributed to dominant surface effect \cite{cab-PRB09}. The results were interpreted in terms of microstructural parameters characterizing the maghemite shells by means of  atomistic Monte Carlo simulations of an individual spherical shell. The model was comprised  of strongly interacting crystallographic domains arranged in a spherical shell with random orientations and anisotropy axis. The Monte Carlo simulation provided the discernment between influence of polycrystalline structure and its hollow geometry, while revealing the magnetic domain arrangement in different temperature regimes. 
 The small EB effect was found in $\alpha-$Fe$_2$O$_3$ nanoleaves when sample was cooled in 20 kOe \cite{xu-PhysicaE09}. This small EB was attributed to  different magnetic order on surface of nanoleaves or coexistence of a minor Fe$_3$O$_4$ phase. The signature of low temperature surface SG layer was also proposed in the high-pressed nanocompacts of Fe$_3$O$_4$ \cite{wang-PRB04}. The magnetic core-shell structure in the single-phase nanoparticle was found to arise during compaction which was evident from field dependence of freezing temperature following the de Almeida–Thouless relationship and the EB effect. Moreover, They found a  critical value of $H_{cool}$, above which both the surface SG behaviour and the EB effect abruptly disappeared. 

\subsubsection{Cobalt oxides}

Nanostructured CoO is historically significant, because EB effect was first discovered by Meiklejohn and Bean \cite{meik1} in a Co/CoO nanoparticles where CoO is typically in a AFM state at low temperature. Recently, EB effect was reported in the monodispersed CoO nanocrystals \cite{zhang-nanotech05,gruy-EPL07,pana-CDG09}. Chemically synthesized CoO nanocrystals having various average particle sizes were investigated where 
FM interaction was appeared at low temperature due to existence of uncompensated moments on surface of the nanoparticles \cite{zhang-nanotech05,pana-CDG09}. The weak FM interaction was found to increase with decreasing particle size. The EB effect was observed for the particles with various sizes synthesized from different chemical routes   which were attributed to coupling between AFM CoO core and surfaces with uncompensated moments. Nanostructured CoO with 1.5 nm thickness grown on Si(111) substrate also showed a strong exchange anisotropy which was attributed to an interaction between AFM order and uncompensated spins in the AFM material \cite{gruy-EPL07}. 

The EB effect has been observed in nanostructured Co$_3$O$_4$ prepared from different routes having different structures viz., nanoparticles, nanowires \cite{sala-nano-lett06,beni-PRL08,dutt-JPCM08,zhu-PhysicaB08,ozka-Eur-J-chem09,beni-EPL09}. Nearly similar EB effect was demonstrated in nanowire prepared from nanocasting  \cite{sala-nano-lett06} and chemical routes \cite{beni-PRL08}. In addition to the shift of magnetic hysteresis loops due to field cooling, training effect further supported the EB effect in nanowire of Co$_3$O$_4$. Temperature dependence of $H_E$ indicated that it vanished close to maximum, observed in the ZFC magnetization. From the anomalous $H_{cool}$ dependence of $H_E$ it was suggested that uncompensated spins at the surfaces were aligned with the increase in $H_{cool}$. Magnetic properties of nanocrystalline Co$_3$O$_4$ was compared with the bulk counterpart where EB effect associated with a considerable $H_E$ (350 Oe) and $H_C$ (250 Oe) were noticed along with the training effect for the  nanoparticles \cite{dutt-JPCM08}. 

\subsubsection{AB$_2$O$_4$-type oxides with spinel structure} 
The EB effect has been reported in few oxides,  Ni$_{0.25}$Co$_{0.25}$Zn$_{0.5}$Fe$_2$O$_4$ \cite{lak-PRB09}, CoFe$_2$O$_4$ \cite{mum-JMMM07},  CoCr$_2$O$_4$ \cite{li-JMMM06} having spinel structure. The EB effect was investigated on nanocrystalline Ni$_{0.25}$Co$_{0.25}$Zn$_{0.5}$Fe$_2$O$_4$ using a combination of in-field, low-temperature M\"{o}ssbauer spectroscopy and dc magnetization studies \cite{lak-PRB09}. The shifts of magnetic hysteresis loop as a function of cooling was studied to characterize EB effect.  In-field M\"{o}ssbauer spectroscopy clearly established coexistence of core and shell contributions which confirmed that 70 \% of spins are in the shell. The barrier energy has been estimated to be 17$\times$10$^{-14}$ ergs which indicated that the shell was not affected by application of large field even at 50 kOe. The system could be modeled as an ordered core with conventional collinear arrangement of spins at the $A$ and $B$ sites and a canted highly frustrated surface. The fairly strong EB was attributed to exchange coupling between core and shell magnetic structures.

\subsubsection{Manganites with perovskite structure}

In last two years EB effect attributed to surface effect has been reported extensively in nanocrystalline mixed-valent manganites with perovskite structure. Those are CaMnO$_3$ \cite{mark-PRB08}, La$_{0.25}$Ca$_{0.75}$MnO$_3$ \cite{huang-PRB08} and La$_{0.2}$Ca$_{0.8}$MnO$_3$ \cite{mark10},  Pr$_{0.5}$Ca$_{0.5}$MnO$_3$ \cite{zhang-PRB09}, Nd$_{0.5}$Ca$_{0.5}$MnO$_3$  \cite{liu-NCMO-JPD}, Sm$_{0.5}$Ca$_{0.5}$MnO$_3$ \cite{zhou-APL08}, La$_{1-x}$Sr$_x$MnO$_3$ \cite{tian-CM06} and double perovskite,  Sr$_2$FeMoO$_6$ \cite{sugata1}.   

The EB effect was observed in compacted CaMnO$_{3-\delta}$ nanoparticles having average particle size $\approx$ 50 nm. Asymmetric hysteresis loops were observed for the measurements in between $\pm$ 90 kOe after field cooling at low temperature. The EB was attributed to the coupling between AFM core and FM grain boundary region. They further suggested that high $H_C$ and EB effect were ascribed to dominating surface effect. The EB was observed in La$_{0.25}$Ca$_{0.75}$MnO$_3$ nanoparticles with average sizes ranging from 40 nm to 1000 nm \cite{huang-PRB08}. The variations of $H_E$ and $H_C$ at $T$ = 5 K with particle size was found to follow nonmonotonic dependencies, showing a  maximum for particles with average diameter around $\sim$ 80 nm. The EB effect was suggested due to uncompensated surface spins of the nanoparticle. They further demonstrated linear relationship between the horizontal and vertical shifts, suggesting that the characteristics of uncompensated spins play an important role on the EB phenomenon. Similar observation of EB effect attributed to AFM core and FM shell structure was recently reported for nanoparticles with slightly different composition, La$_{0.2}$Ca$_{0.8}$MnO$_3$ where evidence of FM component suggested at the grain boundary was observed by the sharp rise in magnetization at low temperature \cite{mark10}.  
The grain-size effects on charge ordering and EB were investigated in a CE-type AFM charge ordered manganite, Pr$_{0.5}$Ca$_{0.5}$MnO$_3$ \cite{zhang-PRB09}. With decreasing size antiferromagnetism as well as charge ordering was found to suppress and charge ordering was absent for size smaller than 40 nm. Suppression of charge ordering and appearance of ferromagnetism were found in nanocrystalline, Nd$_{0.5}$Ca$_{0.5}$MnO$_3$ where dominating surface effect was attributed to the EB phenomenon \cite{liu-NCMO-JPD}. Strong grain size effect of EB was also observed in another charge ordered manganite, Sm$_{0.5}$Ca$_{0.5}$Mn$_3$ attributed to the surface effect \cite{zhou-APL08}. The $H_E$  showed a nonmonotonic variation with average grain size having a maximum value at $\sim$ 120 nm. The EB effect has been observed in  nanoparticle of La$_{1-x}$Sr$_x$MnO$_3$ (LSMO) having average size $\sim$ 20 nm \cite{tian-CM06}. The EB effect was noticed with compositions at $x$ = 0.3, 0.5 and 0.7 where $H_E$ was found to increase monotonously with $x$ from 25 Oe to 250 Oe. It has been interpreted that LSMO nanoparticles had an organic shell, leading to their surface layer being different from the interior and induced spin disorder at the particle surface. 

A weak EB effect attributed to pinning at FM-core and disordered magnetic-shell structure was reported in nanocrystalline Sr$_2$FeMoO$_6$ \cite{sugata1}. The authors suggested that this results provided an important clue for interpreting unusual
tunneling magnetoresistance responses in nanocrystalline Sr$_2$FeMoO$_6$. 
So far EB effect in manganites discussed here are limited to the nanocrystalline materials where surface to bulk ratio is typically enhanced by reducing average grain size. The EB effect can also be observed even in a bulk polycrystalline compound attributed to magnetic core-shell structure. The evidence of EB effect has also been observed in sol-gel derived  polycrystalline, La$_2$NiMnO$_6$ compounds \cite{wang-APL09}. The observed EB was suggested to  originate from coupling between FM La$_2$NiMnO$_6$ and AFM antiphase boundaries. The evidence of antiphase boundaries is one of the crucial issues in double perovskite which has been confirmed through investigation of the EB effect.

\subsubsection{La$_{1/3}$Sr$_{2/3}$FeO$_3$}

\begin{figure}[t]
\centering
\includegraphics[width = 8 cm]{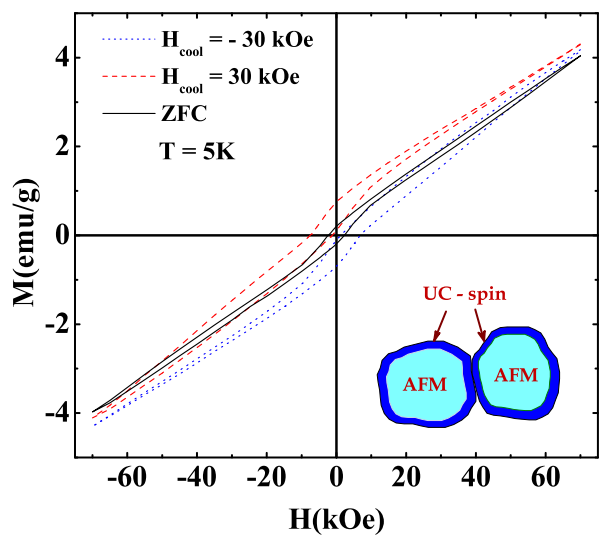}
\caption {Magnetic hysteresis loop measured at 5 K for nanocrystalline  La$_{1/3}$Sr$_{2/3}$FeO$_3$ with average grain size $\sim$ 60 nm where broken curve displays the shifted loops after cooling in $\pm$ 30 kOe and continuous curve displays the loop after cooling in zero-field. Inset exhibits the cartoon of the possible magnetic surface phase separation scenario \cite{sabyasachi}.}
\end{figure}

The strong EB effect is recently observed in another charge ordered compound with composition, La$_{1/3}$Sr$_{2/3}$FeO$_3$ \cite{sabyasachi}. The bulk counterpart does not show any EB effect. However, very strong EB effect is observed in the compound having  $\sim$ 60 nm average grain size. The shifted magnetic hysteresis loops at 5 K due to field cooling at $\pm$ 30 kOe are displayed in figure 14 by the broken curves. The shift is positive for $H_{cool}$ = -30 kOe and +30 kOe which are typical manifestation of the EB effect. Training effect was observed. From the horizontal shift of hysteresis loop $H_E$ was estimated $\sim$ 5.0 kOe at 5 K for $H_{cool}$ at 5 kOe which was monotonically increased to $\sim$ 9.0 kOe for $H_{cool}$ = 50 kOe. It was noted that AFM component dominates in the magnetic hysteresis loop associated with small FM component having considerable coercivity $\approx$ 2.5 kOe. Uncompensated spins at the surface of grains was suggested to interpret FM component where  pinning effect between FM-surface and AFM-core leads to the EB effect in La$_{1/3}$Sr$_{2/3}$FeO$_3$. The proposed magnetic core-shell structure is shown by the carton in the inset of figure 14.


\subsection{Concluding remarks}
The EB effect attributed to magnetic core-shell structure in a structurally single phase compound is much reported in oxides compared to few examples in alloys. However, occurrence of strong surface anisotropy in nanocrystalline alloys and compounds can be confirmed by the EB effect. The issues of finite-size and surface effects in nanocrystalline compounds are currently of much interest to the community. The investigation  of EB effect associated with its cooling field and temperature dependences partially settles the issues of surface anisotropy even from the macroscopic experimental results such as magnetization studies. Still experimental results and theoretical interpretation are not adequately available which needs to be explored comprehensively. 

\section{Exchange bias effects seen in magnetoresistance data}

\subsection{Magnetic field dependence}

\begin{figure}[t]
\centering
\includegraphics[width = 11 cm]{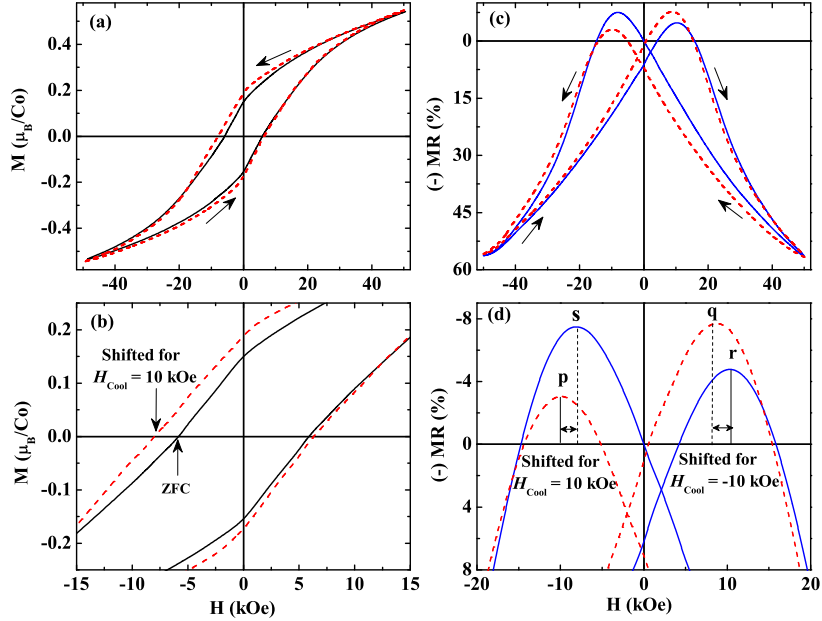}
\caption {Magnetic hysteresis loops (a) for $H_{cool}$ = 0 and 10 kOe (broken curves) and MR-$H$ curves (c) for $H_{cool}$ = 10 (broken curves) and -10 kOe for the polycrystalline  La$_{0.88}$Sr$_{0.12}$CoO$_3$. Central part of the hysteresis loops and MR-$H$ curves are highlighted in (b) and (d), respectively [reprinted from \cite{patraEPL}].}
\end{figure}

The studies of magnetoresistance have been traditionally used to investigate spin valve or giant magnetoresistive or colossal magnetoresistive effect. There are few reports of using MR to study the EB effect. Leighton {\it et al}. carefully investigated EB using MR in Fe films exchange coupled to AFM MnF$_2$ layers where EB phenomenon and $H_C$  obtained from MR were in close agreement with magnetization result. \cite{leigh_MR}. Recently, Bobo {\it et al}. described few examples of EB using MR in a review \cite{bobo}. The studies of EB using MR has also been observed in very few structurally single-phase compounds    \cite{patraEPL,patraJPCM2,patra-EB-JPCM10,xu-MR-EB-JPC,will-PRB09}. First experimental evidence was reported by Patra {\it et al}. in polycrystalline La$_{1-x}$Sr$_x$CoO$_3$ at $x$ = 0.12 close to the percolation threshold of conductivity \cite{patraEPL}. When the  sample was cooled in 10 kOe the shift of magnetic hysteresis loop along the field axis was observed at 5 K which is shown in figures 15(a) and 15(b). Similarly the positive and negative shifts were observed in MR-$H$ curve for negative and positive cooling field (10 kOe), respectively which are shown in figures 15(c) and 15(d). Furthermore, the MR-$H$ measured from -50 kOe to 50 kOe was changed for positive $H_{cool}$ whereas MR-$H$ curve measured from 50 kOe to -50 kOe was changed for negative $H_{cool}$ which are displayed in figure 15(c). Interestingly, $H_E$ estimated from the horizontal shift along the field axis was found to be much higher for the MR-$H$ curve ($\sim$ 940 Oe for $H_{cool}$ = 10 kOe) than that of the estimate from magnetic hysteresis loop ($\sim$ 650 Oe for $H_{cool}$ = 10 kOe). It was suggested that larger EB effect was involved with $H_C$ in accordance with the intuitive model \cite{meik3} and sophisticated  theories \cite{wein-PRB07,hu-EPJB04}. However, the values of $H_{C}/H_{E}$ were nearly close which are $\approx$ 11.2 \% and 10.7 \% obtained from magnetic hysteresis loop and MR-$H$ curve, respectively. Interestingly, the training effect was convincingly observed in the shift of MR-$H$ curve. It was noted that equation (1) for $\lambda \geq$ 2 and equation (2) satisfactorily explain the training effect in  La$_{0.88}$Sr$_{0.12}$CoO$_3$. The EB in polycrystalline, La$_{0.88}$Sr$_{0.12}$CoO$_3$ was attributed to the spontaneous phase separation between FM and SG regions, analogous to that found in Nd$_{0.84}$Sr$_{0.16}$CoO$_3$ given in the insets of figure 16. Similar to that observed in La$_{1-x}$Sr$_x$CoO$_3$, the EB effect in MR-$H$ curve and magnetic hysteresis loop was observed in polycrystalline, Nd$_{0.84}$Sr$_{0.16}$CoO$_3$ close to the percolation threshold of conductivity. This was also attributed to the spontaneous phase separation and pinning effect at the FM and FIM interface \cite{patraJPCM2}. Large $H_{E}$ associated with large $H_C$ was observed in MR-$H$ curve ($H_E$ $\sim$ 744 Oe) than magnetic hysteresis loop ($H_E$ $\sim$ 250 Oe). Training effect was also noticed for Nd$_{0.84}$Sr$_{0.16}$CoO$_3$ which is in accordance with equation (1) for $\lambda \geq$ 2 and equation (2) which is demonstrated in figure 3. A signature of the shift along field axis due to FC process has been reported in polycrystalline double perovskite  compounds, Ca$_2$FeMoO$_6$, Ba$_2$FeMoO$_6$ and Sr$_2$FeReO$_6$ \cite{sugata2}. This was proposed to be a manifestation of EB phenomenon. The FM-core and disordered magnetic-shell structure was proposed to interpret the EB effect. The results on EB effect partly settled the issue of unusual MR in these double perovskite compounds. 

\subsection{Time dependence}

\begin{figure}[t]
\centering
\includegraphics[width = 8 cm]{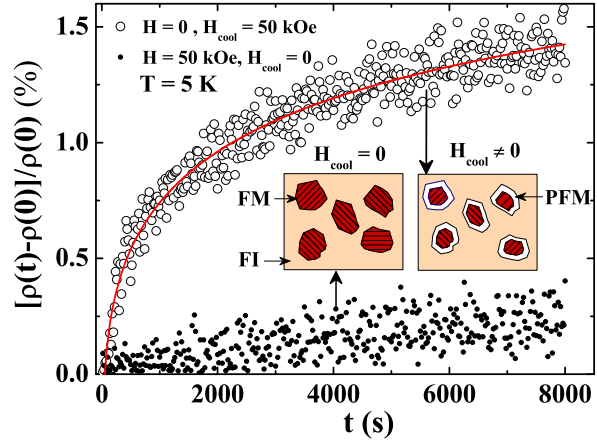}
\caption {Time ($t$) evolution of $[\rho(t)-\rho(0)]/\rho(0)$ at 5 K measured in zero field (open symbols) and $H$ = 50 kOe (filled symbols) after cooling in FC and ZFC modes, respectively in Nd$_{0.84}$Sr$_{0.16}$CoO$_3$. Inset shows diagrams of the grain interior magnetic nanostructure. FM, PFM and FIM represent the ferromagnetic, pinned ferromagnetic  and ferrimagnetic regions [reprinted from \cite{patraJPCM2}].}
\end{figure}

Interestingly, a considerably large time dependence of $\rho(t)$ was observed at 5 K which was involved with the EB effect due to FC process. The value of $[\rho(t)-\rho(0)]/\rho(0)$ at 5 K was $\approx$ 1.4 \% for Nd$_{0.84}$Sr$_{0.16}$CoO$_3$ \cite{patraJPCM2}. In case of FC measurement sample was cooled down to 5 K in FC mode ($H_{cool}$ = 50 kOe) and time dependence was recorded just after switching off the field after stabilizing temperature at 5 K. On the other hand, sample was cooled in ZFC mode down to 5 K and time dependence was recorded in 50 kOe at 5 K. In the second measurement time dependence was found negligible. The results are displayed in figure 16. Similar results were also observed in La$_{0.88}$Sr$_{0.12}$CoO$_3$ with significantly high value of $[\rho(t)-\rho(0)]/\rho(0)$ $\approx$ 10.4 \% at 5 K due to FC process \cite{patraEPL}. Time dependence for both the oxides could be analysed by the stretched exponential 
\begin{equation}
[\rho(t)-\rho(0)]/\rho(0) = A + B \exp(t/\tau)^{\beta} 
\end{equation}  
which has been typically used to analyse time dependence of magnetization for SG compound \cite{mydosh}. The relaxation follows the stretched exponential with
a critical exponent, 0 $< \beta <$ 1 when it is involved with activation against distribution of anisotropy barriers typically observed in the compound, exhibiting glassy magnetic behaviour. The relaxation time ($\tau$) and $\beta$ were found to be 4374 s and 
0.33 for Nd-compound and 2435 s and 0.94 for La-compound, respectively. An example of the fit of time dependence of resistivity measured after field cooled condition is shown in figure 16 by the continuous curve for Nd$_{0.84}$Sr$_{0.16}$CoO$_3$. It is interesting to note that glassy magnetic behaviour of the PFM component was noticed even in a spontaneously phase separated compound, Nd$_{0.84}$Sr$_{0.16}$CoO$_3$ where FM and FIM phases coexist. The signature of EB in time dependence of resistivity is observed only in case of FC measurements for both the compounds. Appearance of the frozen or pinned FM layer at the  interface between FM/SG or FM/FIM components has been proposed to interpret considerable time dependence in resistivity. The term 'pinned' or 'frozen' was used for pinning of FM spins at the interface by hard magnetic component composed of either FIM or SG phase. It was further noted that both the pinned FM and frozen FM layer in Nd$_{0.84}$Sr$_{0.16}$CoO$_3$ and La$_{0.88}$Sr$_{0.12}$CoO$_3$ could be fitted with the stretched exponential function which has been used to interpret relaxation of magnetization for glassy magnetic compounds. This opens up an issue to understand the properties of pinned or frozen FM layers for variety of materials. 

The existence of PFM layer at the interface due to field cooling has been proposed for Nd$_{0.84}$Sr$_{0.16}$CoO$_3$ in the insets of figure 16 which has also been suggested in La$_{0.88}$Sr$_{0.12}$CoO$_3$ \cite{patraEPL}. In both the cases signature of EB effect was noticed in temperature dependence of resistivity at low temperature. The magnitude of resistivity measured in heating mode after cooling the sample in FC process was considerably higher than that of the value found after cooling the sample in ZFC mode. This higher magnitude of resistivity in FC mode was suggested due to appearance of the PFM layer. Analogous to these results, signature of EB effect in the temperature dependence of resistivity has also been reported in SrFeO$_x$ \cite{will-PRB09}. 

\subsection{Concluding remarks}

Signature of EB effect in field, temperature and time dependence of resistivity is rarely noticed contrast to numerous reports observed in magnetic hysteresis loops for structurally single-phase compounds. The most interesting point is to be noted that $H_E$  obtained from  the shift in MR-$H$ curves are considerably higher than the values obtained from the shift of magnetic hysteresis loops. Although different measurement techniques were used in the same sample for investigating EB phenomenology, magnetization results provides overall bulk response whereas MR in the above two examples has been interpreted in terms of tunneling mechanism. Tunneling of current takes place between FM clusters through the non-FM matrix where tunneling barrier is set by anisotropy of the matrix (either SG or FIM discussed in this review) component. During FC process tunneling barrier is modified by appearance of a new layer (frozen or pinned ferromagnetic layer) which leads to asymmetry in MR-$H$ curve and thus the shift is observed. Thus, difference in $H_E$ obtained from two different techniques is specific to these cobaltite compounds as discussed in this review. However, fundamental understanding of EB effect through the measurements of MR needs to be improved through the extensive experimental and theoretical investigations.

\section{General concluding remarks}
We have reviewed the EB effect in variety of structurally single-phase alloys and compounds. The magnetism of alloys and compounds having different types of coexisting magnetic phases are discussed based on the results involved with EB effects. The magnetic structures in alloys and compounds where EB is observed, are categorised into spontaneously phase separated systems and materials having magnetic core-shell structures. The parameters involved with the EB effect such as change in composition due to substitution, grain size effect, cooling field dependence, maximum field used for the measurements of hysteresis loop, temperature dependence, time dependence have been discussed focusing the qualitative and quantitative (using simplified models) aspects of magnetism in variety of alloys and compounds. In few cases quantitative estimates are in accordance with the results obtained from the powerful microscopic experimental tool such as neutron studies. The experimental results and the theories in this context are quite adequately reported so far which open up the following fundamental issues to the community.  

The EB phenomenon is an interface effect which has been observed in variety of single-phase systems and more extensively in various heterostructures. It has been noticed in various possible combinations of soft and hard magnetic substances in several kinds of  interfaces that create difficulties for understanding the phenomenon. The phenomenology has been envisaged that pinned or frozen ferromagnetic layer in various possible ways [for example, displayed in figure 1(b) and right inset of figure 16] typically gives rise to the EB effect. The magnetism of this layer with different combinations of magnetic phases needs to be characterised theoretically and experimentally using microscopic experimental tools. 

In different observations discussed here, the large EB effect is found to be involved with large coercivity. Despite the experimental investigations dealing with the role of anisotropy seem to agree with few proposed theories, any quantitative conclusions from them could not be established because of the difficulties for extracting anisotropy of individual component from the magnetic heterostructures. This issue needs to be investigated extensively from both the experimental and theoretical point of views. 

Plenty of reports involved with the EB effect in few alloys and adequate number of oxide  compounds are found in magnetic core-shell nanostructures having FM or AFM or FIM core magnetism. In last few years EB attributed to finite size effect and surface anisotropy has also been observed in different categories of oxides. More extensive investigations on experimental and theoretical understanding of EB phenomenology are required to confirm the  origin of surface anisotropy. In this context, one of the important issues is whether the EB effect attributed to surface anisotropy belongs to the particular class of materials where core magnetism is the leading factor for such surface anisotropy. Or surface spin disorder (or spin canting having uncompensated spins in AFM or FIM compounds) is ascribed to the presence of nonmagnetic (organic or inorganic) substances at the grain boundaries, or types of nanocrystalline sample preparation are important for the observed surface anisotropy. 

It has been understood that EB in single-phase alloys and compounds confirms magnetically inhomogeneous phases, particularly in the spontaneously phase separated oxides. Interestingly, EB effect attributed to the spontaneous phase separation has recently been observed in very few multiferroic materials. One of the promising examples of coexisting multiferroicity and EB has been found in structurally single phased BiFeO$_3$ \cite{all-APL09,leb-PRB10}. Recently, the unique examples of coexisting multiferroicity and EB has been reported in layer structures where AFM Cr$_2$O$_3$ \cite{EB-ME1,EB-ME2} and BiFeO$_3$ \cite{EB-ME3} having electromagnetic coupling were used for pinning mechanism,  giving rise to the EB effect. In case of layered structure composed of Cr$_2$O$_3$ electrical field was interestingly used to manipulate EB effect \cite{EB-ME1,EB-ME2}. Although nature of AFM states are different, $T_N$ commonly observed above room temperature in both the compositions is promising for the technological application. Thus, coexistence of multiferrocity and EB in the single-phase materials needs to be explored extensively for searching new multiferroic materials having tremendous technological applications.

\vspace{0.5cm}
\noindent{\bf Acknowledgments}
\par

We are indebted to J. Nogu\'{e}s for critical reading of the manuscript. We thank him for  useful comments and valuable suggestions to improve this review. Special thank to \'{O}. Iglesias for illuminating discussions and suggestions. We thank P. Nordblad and L. Bergstr\"{o}m for reading the manuscript. Useful helps from A. Karmakar, S. Das, Sk. Sabyasachi and K. De are acknowledged. S.G. wishes to thank R. Ray for constant encouragement and support during writing this review. S.G. also wishes to thank Department of Science and Technology (Project No. SR/S2/CMP-46/2003), India for the financial support.

\end{document}